\documentclass[Afour,sageh,times]{sagej}
\pdfoutput=1
\usepackage{moreverb,url}
\usepackage{multirow}

\usepackage[colorlinks,bookmarksopen,bookmarksnumbered,citecolor=red,urlcolor=red]{hyperref}

\newcommand\BibTeX{{\rmfamily B\kern-.05em \textsc{i\kern-.025em b}\kern-.08em
T\kern-.1667em\lower.7ex\hbox{E}\kern-.125emX}}

\def\volumeyear{2016}

\begin{document}

\runninghead{Rampeltshammer \textit{et~al.}}

\title{Evaluation and comparison of SEA torque controllers in a unified framework}

\author{Wolfgang Rampeltshammer\affilnum{1}, Arvid Keemink \affilnum{1}, Menno Sytsma\affilnum{1}, Edwin van Asseldonk\affilnum{1}, and Herman van der Kooij\affilnum{1,2}}

\affiliation{\affilnum{1} University of Twente, The Netherlands\\
\affilnum{2} TU Delft, The Netherlands}

\corrauth{Wolfgang Rampeltshammer, Biomechanical Engineering, University of Twente,
Horst Complex  W111, P.O. Box 217, 
7500 AE Enschede,NL}

\email{w.f.rampeltshammer@utwente.nl}

\begin{abstract}
Series elastic actuators~(SEA) with their inherent compliance offer a safe torque source for robots that are interacting with various environments, including humans. These applications have high requirements for the SEA torque controllers, both in the torque response as well as interaction behavior with its the environment. To differentiate state of the art torque controllers, this work is introducing a unifying theoretical and experimental framework that compares controllers based on their torque transfer behavior, their apparent impedance behavior, and especially the passivity of the apparent impedance, i.e. their interaction stability, as well as their sensitivity to sensor noise. We compare classical SEA control approaches such as cascaded PID controllers and full state feedback controllers with advanced controllers using disturbance observers, acceleration feedback and adaptation rules. Simulations and experiments demonstrate the trade-off between stable interactions, high bandwidths and low noise levels. Based on these trade-offs, an application specific controller can be designed and tuned, based on desired interaction with the respective environment.
\end{abstract}


\maketitle

\section{Introduction}
The next big step in robotics is to move them outside structured lab environments to an unstructured real world, where interactions with its environment, including collaborating humans or users for exoskeleton, are less predictable. This necessitates actuation and control that prevent both the environment and the robot from being damaged, especially in cases of human-robot-interaction. One such actuation method is Series Elastic Actuators~(SEAs) with their compliant nature. Accordingly there exists a broad set of proposed control laws that try to most effectively utilize these actuators in specific application scenarios, such as humanoid robots~\citep{hopkins2015embedded, radford2015valkyrie, tsagarakis2013compliant} or exoskeletons~\citep{meijneke2021symbitron, wang2014design, witte2015design}. 

Current research into control of these application scenarios is limited to interactions with a finite set of mass-spring-damper systems. However, this research into static interactions ignores transitions between interaction scenarios, such as impacts during walking, or a robot getting in contact with an object. As a result, it is hard to decide on a control law for a SEA, which is used in a complex interaction scenario, because it is unclear, whether it will allow for stable interaction in unknown environments and how well it will work for a given set of sensors. Therefore, this work presents for the first time a structured comparison for SEA control laws with respect to torque tracking accuracy, interaction with a generalized environment and specifically impacts, i.e. the passivity of the controlled actuator, and noise sensitivity. To increase the usefulness of this comparison for practical implementations, theoretical results are verified experimentally. This comparison provides a guideline for selecting a control law that has unconditional stability while interacting with any environment, while being robust enough for the target application and sensor selection.

\subsection{State of the art}
The original controller by~\citet{pratt1995series} utilizes a PID feedback controller on the interaction force with feedforward terms and load side acceleration feedback. The PD feedback controller was further investigated by~\citet{robinson1999series}, who also investigated effects of the torque controller on the mechanical output impedance of the actuator. \citet{calanca2018rationale} further investigated the benefits of acceleration feedback by investigating its effect on torque tracking while interacting with different mass-spring-damper environments. However, such controllers need accurate sensors for force and acceleration feedback. As a result, cascaded PID controllers were introduced by \citet{sensinger2006improvements}, \citet{vallery2007passive}, and \citet{wyeth2008demonstrating} relying on an inner PI feedback controller of the motor velocity and an outer loop PID controller for the interaction torque. \citet{vallery2007passive} proposed a tuning method that ensures interaction stability of the controlled SEA. Again, sensor capabilities limited their use to cascaded PI controller~\citep{vallery2008compliant}.

Similarly, \citet{ott2004passivity} introduced a P controller of the interaction torque with additional positive torque feedback to reduce the effective inertia or increase the bandwidth of SEAs. This approach was extended by \citet{albu2007unified} by additionally injecting damping with feedback of the interaction torque rate thus reducing the overshoot in the torque response of the controlled actuator. The resulting controller constitutes a full state feedback controller of the interaction torque. A similar approach was proposed by \citet{ragonesi2011series} and \citet{losey2016time} who injected damping by adding motor velocity feedback. In general, these control approaches were mostly successfully focused on improving the bandwidth and overshoot of SEA torque controllers, and less on guaranteeing interaction stability.

With improved sensors, i.e. feedback signals with less noise and a resulting decreased impact of the noise sensitivity, more complex controllers were possible. A commonly seen controller is a combination of a PID controller with feedforward term and with an outer loop disturbance observer~(DOB)~\citep{kong2009control, paine2015actuator, kim2017enhancing, rampeltshammer2020improved} or an inner loop DOB~\citep{kong2013nominal, hopkins2015embedded, haninger2020safe}. \citet{kong2011compact} used a variation of a disturbance observer without a nominal model. \citet{oh2016high} included the immediate load side dynamics in the DOB for better tracking accuracy. \citet{asignacion2021high} used a DOB variation alongside a cascaded PD controller, similar to~\citet{vallery2007passive}. The addition of the DOB improves the torque tracking by removing load side dynamic effects~\citep{paine2015actuator} and load side disturbances~\citep{kong2011compact}, i.e. by reducing the apparent impedance~(also commonly called output impedance) of the actuator~\citep{rampeltshammer2020improved}. However, while the inclusion of DOBs is improving torque tracking, it can also lead to unstable meachnical interactions~\citep{rampeltshammer2020improved}.

Another common approach is the use of adaptive controllers to compensate for regular disturbances or interaction patterns. \citet{calanca2018understanding} proposed the use of a model reference adaptive controller~(MRAC) that adapts to its respective interaction environment by effectively lowering the apparent impedance of the controlled SEA, thus improving torque tracking accuracy in dynamic interaction scenarios. \citet{lin2019decoupled} proposed an adaptive reference controller~(ARC) that learns parameters of a set of unknown non-linear disturbances, and uses those learned parameters to compensate for those disturbances. \citet{zhang2015experimental} proposed a set of controllers that learned a repetitive motion pattern of the output side to improve torque tracking accuracy. These controllers outperform non-adaptive controllers, when a constant interaction environment or a repetitive disturbance, e.g. human motion during locomotion with an exoskeleton, is present. However, their performance advantage vanishes if the interaction is non-repetitive or unpredictable. 

Overall, a broad set of torque controllers exists in literature and has been extensively evaluated for their torque tracking accuracy. However, interaction behavior and stability/passivity have not been analyzed in a unifying framework. A common test for SEAs is to evaluate the torque tracking performance for a regular external disturbance, while the target torque is set to zero~(minimal or zero impedance), e.g. by~\citet{kong2009control} and \citet{paine2015actuator}. Another common method is to test interactions with a specific set of linear environments in the form of mass-spring-damper systems, as done by \citet{paine2015actuator} or \citet{calanca2018understanding}. While the latter method tests a broader set of interactions, it is still limited to a finite set of environments and ignores non-linear interactions such as impacts. Hence, in previous work~\citep{rampeltshammer2020improved} we analyzed the apparent impedance to evaluate the controller performance, and more importantly to ensure that the apparent impedance is passive, i.e. to guarantee unconditional interaction stability. 

\subsection{Our approach}
Current state of the art controllers are either developed directly for a specific application~\citep{ott2004passivity, paine2015actuator} or for a generic actuator~\citep{calanca2018understanding, oh2016high}. Consequently, controller evaluations are application specific, and differences between those controllers, and possible problems of those controllers, are hard to estimate. Additionally, the existing comparisons~\citep{calanca2018rationale} of proposed controllers to the state of the art are limited to the application of the specific controller. This lack of a structured analysis makes it hard to select a controller for a new application. 

Therefore, in this work we present for the first time a structured analysis of SEA torque controllers using a unified framework. Controllers are evaluated based on locked output torque tracking performance, their disturbance rejection, i.e. magnitude of the apparent impedance, their interaction stability, i.e. passivity of the apparent impedance, as well as their noise sensitivity. With this structured analysis, it is possible to select controllers specifically based on application requirements, such as bandwidth requirements, sensor noise levels~\citep{asignacion2021high}, and interaction types, e.g. low apparent impedance if in direct interaction with humans or unconditional interaction stability in case of uncertain interactions. 

For this comparison, we are comparing the following controllers or variations thereof: full state torque feedback~\citep{albu2007unified}, full state motor position feedback~\citep{losey2016time}, cascaded PID~\citep{vallery2007passive}, PD~\citep{robinson1999series}, MRAC as an example of an adaptive controller~\citep{calanca2018understanding} as well as PD-DOB~\citep{rampeltshammer2020improved}, PD with acceleration feedback~\citep{pratt1995series}, and full state torque feedback with DOB~(novel to our knowledge) as well as with acceleration feedback~\citep{calanca2018understanding}. These controllers are compared theoretically, as well as practically, to identify the effects non-linear disturbances such as static friction or cogging have on these controllers. For the theoretical comparison, each controller is tuned for an identical bandwidth and similar overshoot, as well as for a passive apparent impedance. Subsequently, these controllers are compared based on their torque tracking accuracy, apparent impedance magnitude, and noise sensitivity. For the practical tests, controllers are tuned identically, and evaluated based on their torque tracking accuracy, apparent impedance, and additionally impact behavior to experimentally confirm their passivity.

In Section~\ref{sec:methods}, the selected controllers are presented. This is followed by the theoretical comparison of these controllers in Section~\ref{sec:nominal}. The protocol for the experimental evaluation is presented in Section~\ref{sec:expmethods}. The results of the experimental evaluation are presented in Section~\ref{sec:results} and discussed in Section~\ref{sec:discussion}. In Section~\ref{sec:conclusion} this work is concluded.
\begin{table}[h!]
\small\sf\centering
\caption{List of symbols and abbreviations.}
\label{table:symbols}
 \begin{tabular}{r l} 
 \toprule
 Symbol & Definition \\
 \midrule
 $j_m, b_m$ & motor inertia and damping \\
 $k$ & spring stiffness\\
 $q, \theta$ & motor and load position\\
 $\tau_m, \tau_k$ & motor and interaction torque\\
 $\zeta_n, \zeta_d$ & natural and target damping ratio \\
 $\omega_n, \omega_d$ & natural and target frequency \\
 $\omega_{BW}$ & target bandwidth\\
 $\delta_\zeta$ & damping correction factor\\
 \midrule
 \multicolumn{2}{c}{Feedback signal~$n\in\{\tau_k, \dot{q}, \ddot{\theta}\}$}\\
 \midrule
 $\eta_n$ & additive noise for signal~$n$\\
 $\sigma_n$ & standard deviation of~$\eta_n$\\
 $\tilde{n} = n +\eta_n$ & noisy signal~$n$\\
  $s$ & complex frequency\\
 $H, Z, R, Y, Q, T$ & transfer functions\\
 \midrule
 \multicolumn{2}{l}{Controller~$i\in\{ft:\text{FSFt}, fm:\text{FSFm},$}\\
 \multicolumn{2}{c}{$cp:\text{Cascaded PID}, pd:\text{PD}, mr: \text{MRAC}$}\\
 \multicolumn{2}{c}{$dob:\text{DOB}, fa:\text{Acceleration Feedback}\}$}\\
 \multicolumn{2}{l}{Gain~$x\in\{P:\text{Proportional}, D:\text{Derivative},$}\\
 \multicolumn{2}{c}{$I:\text{Integral}, C:\text{Corrective}\}$}\\
 \midrule
 $K_x^{i}$ & controller gain~$x$ for controller~$i$\\
 $C^{i}$ & control law of controller~$i$\\
 $\alpha^{i}\in[0, 1]$ & scaling gain for controller~$i$\\ 
 $H^{i}_c$ & closed loop torque transfer\\
 &  for controller~$n$\\
 $Z^{i}_c$ & closed loop apparent impedance \\
 &for controller~$n$\\
 $~_n\!T^{i}_c$ & closed loop noise sensitivity \\&
 for feedback signal~$n$ of controller~$i$\\
 $N^i_n$ & noise spectral density\\
 &for feedback signal~$n$ of controller~$i$\\
 \bottomrule
 \end{tabular}
\end{table}
\section{Methods}\label{sec:methods}
\begin{figure}
	\centering
	\includegraphics[]{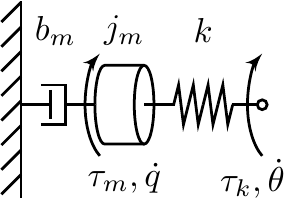}
	\includegraphics[]{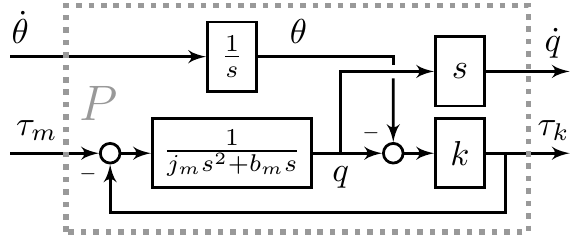}
	\caption{Shown on the top is the physical representation of the SEA, with motor side damping $b_m$, motor inertia $j_m$, spring stiffness $k$, motor torque $\tau_m$ and position $q$, as well as output position $\theta$ and interaction torque $\tau_k$. Additionally, the transfer model of the presented SEA is shown on the bottom.}
	\label{fig:sea}
\end{figure}

All symbols and notation, used in the following sections are described in Table~\ref{table:symbols}. Additionally, they are introduced were first presented. For the purposes of this paper, SEAs are modeled as a rotational joint that is driven by a motor with reflected inertia~$j_m$, has a motor side damping~$b_m$, and interacts with a rotational spring with stiffness~$k$, as shown in Fig.~\ref{fig:sea}. It exhibits the following dynamics:
\begin{equation}
    j_m\ddot{q} = \tau_m-b_m\dot{q}-k\left(q-\theta\right),
\end{equation}
with $q$ denoting the motor position, and $\theta$ denoting the output position of the SEA. The difference between both angles describes the interaction or spring torque~$\tau_k = k\left(q-\theta\right)$. This results in the following transfer behavior:
\begin{align}
    \tau_k & = H\tau_m + Z\dot{\theta} \\\nonumber
    \dot{q} &= R\tau_m + Y\dot{\theta},
\end{align}
with torque transfer~$H$, and apparent impedance~$Z$
\begin{align}
    H &= \frac{\tau_k}{\tau_m} = \frac{k}{j_ms^2+b_ms+k}=\frac{\omega_n^2}{s^2+2\zeta_n\omega_n+\omega_n^2}\\\nonumber
    Z &= \frac{\tau_k}{-\dot{\theta}} = \frac{k\left(j_ms+b_m\right)}{j_ms^2+b_ms+k} = \frac{k\left(s+2\zeta_n\omega_n\right)}{s^2+2\zeta_n\omega_n+\omega_n^2}\\\nonumber
    R &= \frac{\dot{q}}{\tau_m} = \frac{s}{k} H, \quad Y = \frac{\dot{q}}{-\dot{\theta}} = H.
\end{align}
Here~$\zeta_n$ and~$\omega_n$ denote the natural damping ratio and natural frequency of the uncontrolled SEA.

The apparent impedance, or output impedance, is a rewritten form of the load/disturbance sensitivity. It gives an indication on how well the controlled or uncontrolled system rejects disturbances from the load side. In current SEA state of the art, apparent impedance is mostly presented indirectly: one option presents its effects via changes in torque tracking performance at higher frequencies, for specified environments~\citep{calanca2018rationale,paine2015actuator}. Another option shows its effects as the transparency of the device against external motions at specified frequencies~\citep{kong2011compact}. However, these analytical options are limited by the fact that non-linear disturbances, such as impacts, are ignored. As a result, passivity of the apparent impedance is not considered, and unconditional interaction stability cannot be evaluated. Therefore, in this work, we will assess all presented controllers based on their apparent impedance.

In this work, all non-linear effects such as stiction or gearbox effects such as cogging are ignored, i.e. uncompensated, for the design and analysis of controllers. These effects are present in most high-power actuators, and can be compensated to varying degrees~\citep{calanca2018understanding, losey2016time}. Therefore, in this work, controllers are compared in the worst case scenario, where all non-linear disturbances are uncompensated. This gives an additional indication about the robustness of the investigated controllers.

Furthermore, we assume that the system parameters~$j_m$,~$b_m$, and~$k$ are known or have been identified reasonably well.

Additionally, each sensor used in the control law has their own additive noise~$\eta_i$, which results in the noise sensitivities~$_i\!T = \frac{\tau_k}{\eta_i}$. Therefore, the following measured variables are defined:
\begin{align}
    \tilde{\tau}_k &= \tau_k - \eta_{\tau}\\\nonumber
    \ddot{\tilde{\theta}}&=\ddot{\theta}+\eta_{\ddot{\theta}}\\\nonumber
    \dot{\tilde{q}} &= \dot{q} - \eta_{\dot{q}},
\end{align}
with $\tilde{\tau}_k$ being the measured interaction torque, $\ddot{\tilde{\theta}}$ the  measured joint acceleration, and $\dot{\tilde{q}}$ the measured motor velocity. Signs for each noise source were chosen such that the noise sensitivity is positive to simplify notation and readability.

Given that the used control laws result in dissimilar torque transfer behavior, the controllers are tuned based on nominal torque transfer bandwidth $\omega_{BW}$ (defined as the $-3$~dB crossing), and the overshoot of the system, which we define as a target damping ratio~$\zeta_d$, where possible. As such the controlled system is expected to approximate the torque transfer of a controlled mass-spring-damper system with a controlled natural frequency~$\omega_d$.

\subsection{Shaping the torque transfer}
In the following control laws as well as design methods are presented for the full state torque feedback controller, for the full state motor position feedback controller, for the cascaded PID controller, for the PD controller as well as for the MRAC.
\subsubsection{Full state feedback}
A common control method used for series elastic actuators is full state feedback~(FSFB). Ott et al.~\citet{ott2004passivity} used full state feedback to render an impedance, while Losey et al.~\citet{losey2016time} used full state feedback on the motor position for an impedance controller. Here we will compare two FSF options: FSF for the interaction torque~$\tau_k$, which is denoted FSFt, and of the motor position~$q$, as presented by Losey et al.~\cite{losey2016time}, which is denoted FSFm.

\begin{figure}
	\centering
    \includegraphics[]{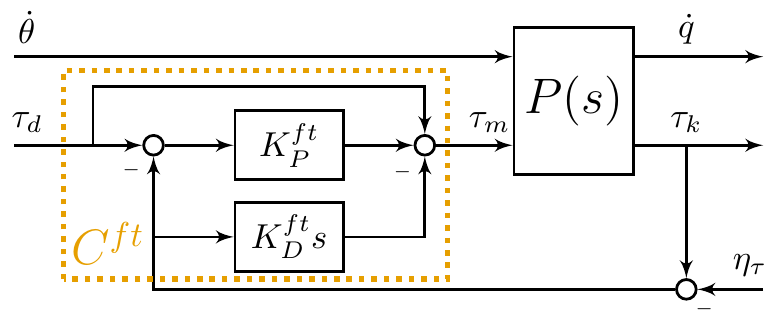}
	\caption{Diagram for the FSFt controller. Controller includes a feedforward term as well as the state feedbacks of the interaction torque $\tau_k$.}
	\label{fig:fsftor}
\end{figure}

For FSFt, as shown in Fig.~\ref{fig:fsftor}, the control law is given as:
\begin{align}
    \tau_m &= \tau_d + K_P^{ft}\left(\tau_d-\tilde{\tau}_k\right) - K_D^{ft}\dot{\tilde{\tau}}_k
\end{align}
which results in
\begin{equation}
    \tau_k = H_c^{ft}\tau_d + Z_c^{ft}\dot{\theta} +  ~_\tau\!T_c^{ft} \eta_{\tau},
\end{equation}
with
\begin{align}
    H_c^{ft} &= \frac{k\left(1+K_P^{ft}\right)}{j_ms^2+\left(b_m+kK_D^{ft}\right)s+ k(1+K_P^{ft})},\\\nonumber
    Z_c^{ft} &= \frac{k\left(j_ms+b_m\right)}{j_ms^2+\left(b_m+kK_D^{ft}\right)s+ k(1+K_P^{ft})},\\\nonumber
    ~_\tau\!T_c^{ft} &= \frac{k\left(K_P^{ft}+K_D^{ft}s\right)}{j_ms^2+\left(b_m+kK_D^{ft}\right)s+ k(1+K_P^{ft})}.
\end{align}
The controller gains are defined as:
\begin{align}\label{eq:fsft_gains}
    K_P^{ft} &= \frac{\omega_d^2}{\omega_n^2}-1\\\nonumber
    K_D^{ft} &= 2\frac{\zeta_d\omega_d-\zeta_n\omega_n}{\omega_n^2}
\end{align}
with
\begin{equation}\label{eq:omega_fs}
    \omega_d = \frac{\omega_{BW}}{\sqrt{1-2\zeta_d^2+\sqrt{1+\left(2\zeta_d^2-1\right)^2}}}
\end{equation}

Another option of full state feedback is to define the motor position as the state, as proposed by~\citet{losey2016time}. It has to be noted that the proposed torque controller is used alongside an impedance controller. As stated in the assumptions, the friction compensation terms and the adaptive learning of the system's parameters is ignored. The resulting rewritten control law for FSFm, as shown in Fig.~\ref{fig:fsfmot}, is defined as follows, as derived in Appendix~\ref{sec:apdx_1}:
\begin{equation}
    \tau_m = \tau_d + K_P^{fm}\left(\tau_d-\tilde{\tau}_k\right) - K_D^{fm}\dot{\tilde{q}}
\end{equation}
which results in
\begin{equation}
    \tau_k = H_{c}^{fm}\tau_{d} + Z_{c}^{fm}\dot{\theta} +  ~_{\tau}\!T_{c}^{fm} \eta_{\tau}+~_{\dot{q}}\!T_{c}^{fm} \eta_{\dot{q}},
\end{equation}
with
\begin{align}
    H_c^{fm} &= \frac{k\left(1+K_P^{fm}\right)}{j_ms^2+\left(b_m+K_D^{fm}\right)s+ k(1+K_P^{fm})},\\\nonumber
    Z_c^{fm} &= \frac{k\left(j_ms+b_m+K_D^{fm}\right)}{j_ms^2+\left(b_m+K_D^{fm}\right)s+ k(1+K_P^{fm})},\\\nonumber
    ~_{\dot{q}}\!T_c^{fm} &= \frac{k\left(K_D^{fm}\right)}{j_ms^2+\left(b_m+K_D^{fm}\right)s+ k(1+K_P^{fm})},\\\nonumber
    ~_\tau\!T_c^{fm} &= \frac{k\left(K_P^{fm}\right)}{j_ms^2+\left(b_m+K_D^{fm}\right)s+ k(1+K_P^{fm})}
\end{align}
The controller gains are defined as
\begin{align}
    K_P^{fm} &= \frac{\omega_d^2}{\omega_n^2}-1\\\nonumber
    K_D^{fm} &= 2j_m\left(\zeta_d\omega_d-\zeta_n\omega_n\right)
\end{align}
with $\omega_d$ as defined in Eq.~\ref{eq:omega_fs}. The difference between FSFt and FSFm is visible in the apparent impedance~$Z_c^{fm}$, where the damping term of the controller is included in the numerator, which as a result increases the apparent impedance at low frequencies. This detrimental effect can be linked to the missing information, if compared to FSFt: Motor position feedback is missing information of the load side, thus limiting the controllers capability to reject errors introduced from the load side.

\begin{figure}
	\centering
    \includegraphics[]{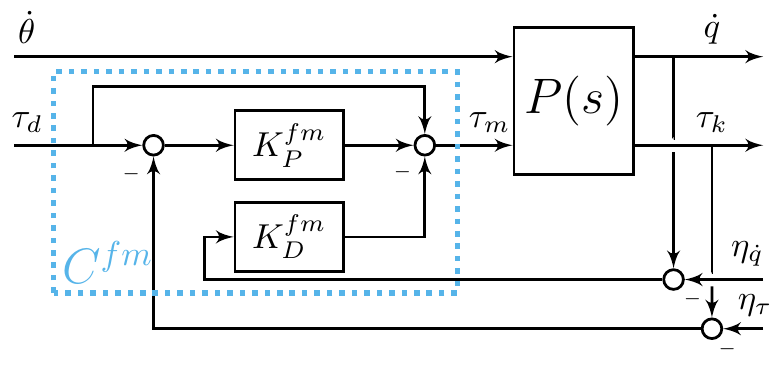}
	\caption{Diagram for the FSFm controller in its adapted form, including a feedforward term and state feedback of the motor velocity $\dot{q}$.}
	\label{fig:fsfmot}
\end{figure}

\subsubsection{Cascaded PID}
\begin{figure}
	\centering
    \includegraphics[]{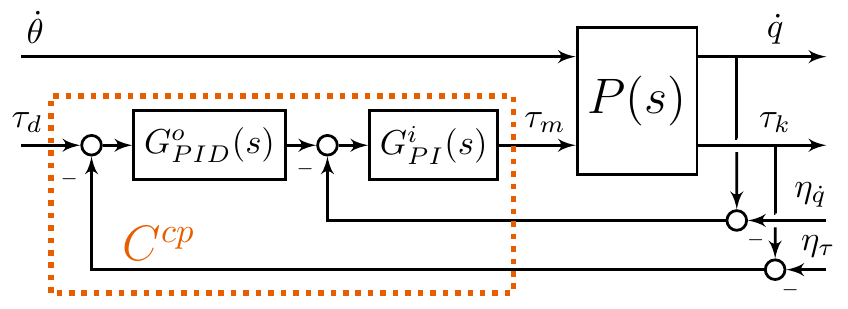}
	\caption{Control diagram of the cascaded PID controller, with the inner velocity loop and its PI controller~$G_{PI}^{i}$, and the outer torque loop and its PID controller~$G_{PID}^{o}$.}
	\label{fig:cascaded}
\end{figure}

Another common approach for controlling SEAs are cascaded PID controllers~\citep{vallery2008compliant, vallery2007passive, sensinger2006improvements}.
\citet{vallery2007passive} proposed an inner loop PI controller of motor velocity with an outer loop PID controller of joint torque and derived conservative passivity limits to ensure a passive apparent impedance. In this work, the outer loop PID approach~\citep{vallery2007passive}, as shown in Fig.~\ref{fig:cascaded}, is used instead of a PI controller~\citep{vallery2008compliant} to retain the option of reducing the overshoot of the torque response. 
\begin{align}
    \tau_m &= G_{PI}^{i}\left(G_{PID}^{o}\left(\tau_d-\tilde{\tau}_k\right)-\dot{\tilde{q}}\right)\\\nonumber
    G_{PI}^{i} &=  K_{P,i}^{cp}+\frac{K_{I,i}^{cp}}{s}\\\nonumber
    G_{PID}^{o} &= K_{P,o}^{cp}+\frac{K_{I,o}^{cp}}{s}+K_{D,o}^{cp}
\end{align}
which results in
\begin{equation}
    \tau_k = H_{c}^{cp}\tau_{d} + Z_{c}^{cp}\dot{\theta} +  ~_{\tau}\!T_{c}^{cp} \eta_{\tau}+~_{\dot{q}}\!T_{c}^{cp} \eta_{\dot{q}},
\end{equation}
with,
\begin{align}
    H_c^{cp} &= k\frac{\sum_{i=0}^3 b_is^i}{\sum_{j=0}^4a_js^j},\\\nonumber
    Z_c^{cp} &= k\frac{j_ms^3+\left(b_m+K_{P,i}^{cp}\right)s^2+K_{I,i}^{cp}s}{\sum_{j=0}^4a_js^j},\\\nonumber
    ~_{\dot{q}}\!T_c^{cp} &=k\frac{K_{P,i}^{cp}s^2+K_{I,i}^{cp}s}{\sum_{j=0}^4a_js^j},\\\nonumber
    ~_{\tau}\!T_c^{cp} &= H_c^{cp}.
\end{align}
with,
\begin{align}
    b_3 &= K_{P,i}^{cp}K_{D,o}^{cp}\\\nonumber
    b_2 &= K_{P,i}^{cp}K_{P,o}^{cp}+K_{I,i}^{cp}K_{D,o}^{cp}\\\nonumber
    b_1 &= K_{P,i}^{cp}K_{I,o}^{cp}+K_{I,i}^{cp}K_{P,o}^{cp}\\\nonumber
    b_0 &= K_{I,i}^{cp}K_{I,o}^{cp}
\end{align}
and
\begin{align}
    a_4 &= j_m\\\nonumber
    a_3 &= b_m+K_{P,i}^{cp}\left(1+kK_{D,o}^{cp}\right)\\\nonumber
    a_2 &= K_{I,i}^{cp}+k\left(1+K_{P,i}^{cp}K_{P,o}^{cp}+K_{I,i}^{cp}K_{D,o}^{cp}\right)\\\nonumber
    a_1 &= k\left(K_{P,i}^{cp}K_{I,o}^{cp}+K_{I,i}^{cp}K_{P,o}^{cp}\right)\\\nonumber
    a_0 &= kK_{I,i}^{cp}K_{I,o}^{cp}.
\end{align}

Furthermore, for shaping the loop, the following tuning rules were proposed by ~\citet{vallery2007passive} that ensure passivity of the apparent impedance:
\begin{align}
    K_{P,i}^{cp} & >j_m\\\nonumber
    K_{I,i}^{cp} & < 0.5K_{P,i}^{cp}\\\nonumber
    K_{I,o}^{cp} & < 0.5K_{P,o}^{cp}
\end{align}
The constraint for $K_{D,o}^{cp}$ is ignored because the motor velocity is available as an unfiltered signal, and as such the constraint can be omitted~\citep{vallery2007passive}. Additionally, it has to be noted that the passivity analysis by~\cite{vallery2007passive} is conservative, and could be further relaxed, if the whole inequality for passivity is analyzed for positive realness instead of each polynomial coefficient (similar to the analysis performed by~\citet{rampeltshammer2020improved}). In this work, we will use the constraints proposed by~\citet{vallery2007passive}, since a derivation and presentation of less conservative constraints would exceed the scope of this work. Additionally, in contrast to~\citet{vallery2008compliant} the outer differential gain is retained to further damp the system.

Furthermore, it has to be noted that tuning of the cascaded PID controller is more complicated compared to all other presented approaches for which a target bandwidth and damping ratio can be selected. However, this problem can be ameliorated by shaping the controller using optimization methods. 

\subsubsection{PD controller}
Another common torque controller for SEAs is a PD controller as presented by~\citet{robinson1999series, paine2015actuator, rampeltshammer2020improved}. The PD controller is mostly used alongside a disturbance observer~(DOB), which is introduced in Sec.~\ref{sec:meth_dob}.
\begin{align}
    \tau_m &= \tau_d + K_P^{pd}\left(\tau_d-\tilde{\tau}_k\right) + K_D^{pd}\left(\dot{\tau}_d-\dot{\tilde{\tau}}_k\right)
\end{align}
which results in
\begin{equation}
    \tau_k = H_c^{pd}\tau_r + Z_c^{pd}\dot{\theta} + T_c^{pd} \eta_{\tau},
\end{equation}
with
\begin{align}
    H_c^{pd} &= \frac{k\left(1+K_P^{pd}+K_D^{pd}s\right)}{j_ms^2+\left(b_m+kK_D^{pd}\right)s+ k(1+K_P^{pd})},\\\nonumber
     Z_c^{pd} &= \frac{j_ms+b_m}{j_ms^2+\left(b_m+kK_D^{pd}\right)s+ k(1+K_P^{pd})},\\\nonumber
     T_c^{pd} &= \frac{k\left(K_P^{pd}+K_D^{pd}s\right)}{j_ms^2+\left(b_m+kK_D^{pd}\right)s+ k(1+K_P^{pd})}.\\
\end{align}

\begin{figure}
	\centering
    \includegraphics[]{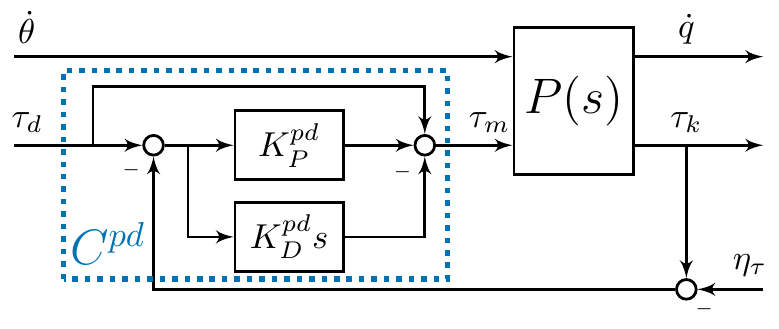}
	\caption{Control diagram of the PD controller with a feedforward term.}
	\label{fig:pd}
\end{figure}
The controller gains are defined identical to FSFt, as described in Eq.~\ref{eq:fsft_gains}. However the controlled frequency has to be amended as described by \citet{rampeltshammer2020improved}:
\begin{align}\label{eq:delta}
    \omega_d &= \frac{\omega_{BW}}{\sqrt{1 - 2\zeta_d^2(1-2\delta_\zeta^2) + \sqrt{1 + (2\zeta_d^2(1-2\delta_\zeta^2) - 1)^2}}},\\\nonumber
    \delta_\zeta &= 1 - \frac{\zeta_n\omega_n}{\zeta_d\omega_c},
\end{align}
with~$\delta_\zeta$ being the damping correction factor.
The adapted equations take the bandwidth increasing effect of the system's zero into consideration and as a result achieve identical bandwidths to FSFt, but with lower gains. The common addition of a DOB is presented in section~\ref{sec:meth_dob} as a method for improving the apparent impedance.

\subsubsection{MRAC}
As an example of adaptive control approaches, we investigate the model reference adaptive controller~(MRAC) presented by~ \citet{calanca2018understanding}. The control law is defined as:
\begin{align}
    \tau_m =& K_P^{mr}\left(\tau_d-\tilde{\tau}_k\right)-K_D^{mr}\dot{\tilde{\tau}}_k + K_C^{mr} H_r^{mr}\dot{\tau}_d\\\nonumber
    & +\hat{b}\dot{\tilde{\tau}}_k + \hat{c}\tilde{\tau}_k
\end{align}
which results in
\begin{equation}
    \tau_k = H_{c}^{mr}\tau_{d} + Z_{c}^{mr}\dot{\theta} +  ~_{\tau}\!T_{c}^{mr} \eta_{\tau},
\end{equation}
if the adaptive components~$\hat{b}$ and~$\hat{c}$ are assumed to have converged. Outside of steady state behavior, linear analysis can never account for all effects.
The transfer functions are defined as:
\begin{align}
    H_r^{mr} &= \frac{\tau_r}{\tau_d} = \frac{\omega_d^2}{s^2+2\zeta_d\omega_d+\omega_d^2},\\\nonumber
    H_c^{mr} &= H_r^{mr}\frac{s^2+2\omega_d s+ \omega_d^2}{s^2 + \left(2\omega_d-\hat{b}\right)s + \left(1-\hat{c}\right)\omega_n+\omega_d^2},\\\nonumber
    Z_c^{mr} &= \frac{\omega_n^2\left(j_ms+b_m\right)}{s^2 + \left(2\omega_d-\hat{b}\right)s + \left(1-\hat{c}\right)\omega_n^2+\omega_d^2}\\\nonumber
    ~_{\tau}\!T_c^{mr} &= \frac{\omega_n^2\left(K_P^{mr}-\hat{c}+K_D^{mr}s-\hat{b}s\right)}{s^2 + \left(2\omega_d-\hat{b}\right)s + \left(1-\hat{c}\right)\omega_n^2+\omega_d^2}.
\end{align}
Here $H_r^{mr}$ is defined as the reference model for the MRAC, and can be tuned based on Eq.~\ref{eq:omega_fs} and a target damping ratio. Additionally, the presented torque transfer~$H_c^{mr}$ is after a pole zero cancellation, in cases of~$\zeta_d\neq 1$. The controller gains are defined as:
\begin{align}
    K_P^{mr} &= \frac{\omega_d^2}{\omega_n^2},\\
    K_D^{mr} &= \frac{2\omega_d-2\zeta_n\omega_n}{\omega_n^2},\\
    K_C^{mr} &=\frac{2\omega_d\left(1-\zeta_d\right)}{\omega_n^2}.
\end{align}
It can be seen that the damping correction term~$K_C^{mr}$ is only contributing, if a damping ratio of $\zeta_d\neq 1$ is selected. For the adaptive gains~$\hat{b}$ and $\hat{c}$, some conclusions can be drawn from looking at the transfer behavior. In case of a locked output, we expect $\hat{b} = 0$ and $\hat{c} = 1$. In all other interaction cases, the role of the adaptive gains can be seen as extending the model such that it eliminates the apparent impedance for the specified output motion. It can be expected that this effect works best for a single excitation frequency, as demonstrated by~\cite{calanca2018understanding}.  The nonlinear behavior of the adaptive gains, as shown below,
\begin{align}
    \dot{\hat{b}} &= -\rho\left(e\dot{\tilde{\tau}}_k+\sigma\hat{b}\right)\\\nonumber
    \dot{\hat{c}} &= -\rho\left(e\tilde{\tau}_k+\sigma\hat{c}\right)\\\nonumber
    e &= \dot{\tilde{\tau}}_k - H_R^{mc}\dot{\tau}_d + \omega_d^2\left(\tilde{\tau}_k - H_r^{mc}\tau_d\right)
\end{align}
will deteriorate controller performance in the presence of multiple excitation frequencies, possibly from mechanical disturbances such as cogging. Additionally to the noise sensitivity of the converged adaptive controller, the non-linear effects of noise on the adaptation also should be considered for a thorough analysis. Such an analysis goes beyond the scope of this work. However, it can be reasoned that the multiplicative effects based on torque and torque rate amplitudes will significantly influence the convergence of these adaptive parameters.
Besides the introduced non-linearity, the controller also introduces learning rates that need to be tuned accordingly. Therefore a trade-off between fast learning and reduced performance at low frequencies, and slow learning and reduced performance at high frequencies needs to be made. This further complicates the tuning process.

\begin{figure}
	\centering
    \includegraphics[]{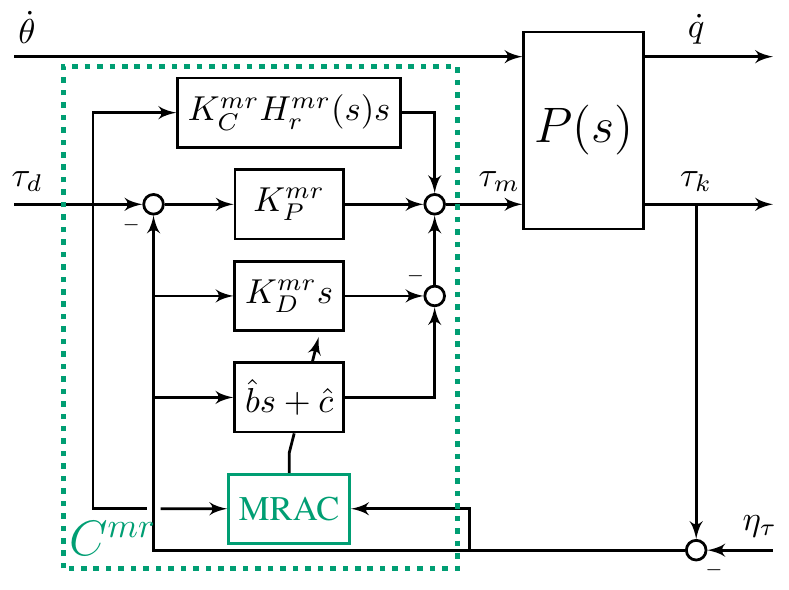}
	\caption{MRAC control scheme including the adaptive control block, and adaptive gains~$\hat{b}$ and~$\hat{c}$.}
	\label{fig:mrac}
\end{figure}

\subsection{Shaping the apparent impedance}
Besides improving the torque response of SEAs, additional components were added to the torque controllers, mainly DOBs~\citep{rampeltshammer2020improved, paine2015actuator} and acceleration feedback~\citep{pratt1995series, calanca2018rationale}. Both methods effectively shape the apparent impedance, by compensating for external motions. Effectively DOBs are better at lowering the apparent impedance at low frequencies~\citep{rampeltshammer2020improved}, while acceleration feedback has a bigger impact at higher frequencies~\citep{calanca2018rationale}. In the following both methods will be introduced and analyzed.

\subsubsection{Disturbance observer}\label{sec:meth_dob}
A common method to compensate for unmodeled disturbances are DOBs. DOBs offer an effective method to compensate for disturbances such as cogging and friction, but need additional consideration for stability and passivity~\citep{schrijver2002disturbance}. In this work, outer loop DOBs are investigated, i.e. it is assumed that the plant is controlled by an arbitrary torque controller~$C^{i}$ that results in a closed loop torque transfer~$H_c^{i}$. Since the purpose of the DOB is to reject disturbances and not to shape the torque loop, it is furthermore assumed that the nominal model is equal to the closed loop torque transfer, i.e.~$H_n = H_c^{i}$.

For such a DOB, the control law is given as:
\begin{equation}
    \tau_r = \left(1-\alpha^{dob} Q^{dob}\right)^{-1}\left(\tau_d - \alpha^{dob} Q^{dob} H_n^{-1}\tau_k\right)
\end{equation}
with $\tau_r$ being the desired torque of the selected inner loop controller, and~$\alpha^{dob}$ and~$Q^{dob}$ respectively being the DOB gain and filter. This results in
\begin{equation}
    \tau_k = H_c^{dob}\tau_d + Z_c^{dob}\dot{\theta} + ~_{\tau}\!T_c^{dob} \eta_{\tau},
\end{equation}
with
\begin{align}
    H_c^{dob} &= H_c^{i}\\\nonumber
    Z_c^{dob} &= \left(1-\alpha^{dob} Q^{dob}\right)Z_c^{i}\\\nonumber
     ~_{\tau}\!T_c^{dob} &= \alpha^{dob} Q^{dob} + \left(1-\alpha^{dob} Q^{dob}\right)~_{\tau}\!T_c^{i}.
\end{align}
The DOB reduces the apparent impedance at low frequencies at the cost of increased phase lead. \citet{rampeltshammer2020improved} demonstrated that this can sometimes lead to the loss of passivity. Therefore, they introduced the DOB gain~$\alpha$ to guarantee a passive apparent impedance.

The DOB filter~$Q$ and DOB gain~$\alpha^{dob}$ have to be selected and tuned based on the selected inner loop torque controller. For an inner loop PD controller or FSFt, the method previously presented by~\citet{rampeltshammer2020improved} can be utilized to find a DOB implementation with a passive apparent impedance. Hence in this work, $Q$ and $\alpha$ are defined as follows:
\begin{align}\label{eq:butter}
    Q^{dob} &=\frac{(\omega_q^{dob})^2}{s^2+\sqrt{2}\omega^{dob}_qs+(\omega_q^{dob})^2}
\end{align}

\begin{figure}
	\centering
    \includegraphics[]{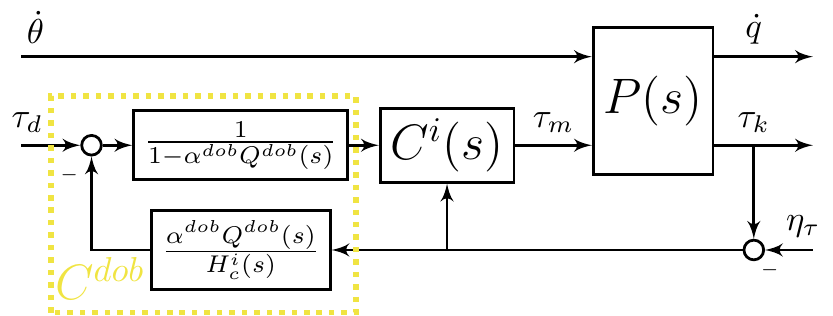}
	\caption{Control scheme for a DOB, for an arbitrary controller $C^{i}$.}
	\label{fig:dob}
\end{figure}
 
\subsubsection{Acceleration feedback}
A separate option to lower the apparent impedance of each system is the introduction of output side acceleration feedback~\citet{pratt1995series, calanca2018rationale}. As already stated and demonstrated by~\citet{calanca2018rationale}, output side acceleration feedback can be combined with any controller~$C^{i}$. For comparison purposes, this acceleration feedback is analyzed in combination with a PD controller and FSFt controller. The general control law is given as:
\begin{equation}
    \tau_m = C^{i}\left(s,\tau_d, \tau_k\right) + \alpha^{fa} j_m Q^{fa}(s)\ddot{\theta}
\end{equation}
with~$\alpha^{fa}$ and $Q^{fa}$ being the acceleration feedback gain and filter respectively. This results in
\begin{equation}
    \tau_k = H_c^{fa}\tau_d + Z_c^{fa}\dot{\theta} +  ~_\tau\!T_c^{fa} \eta_{\tau} +  ~_{\ddot{q}}\!T_c^{fa} \eta_{\ddot{q}},
\end{equation}
with
\begin{align}
    H_c^{fa}(s) &= H_c^{i}(s),\\\nonumber
    Z_c^{fa}(s) &= \frac{k\left(\left(1-\alpha^{fa} Q^{fa}(s)\right)j_ms+b_m\right)} {j_ms^2+\left(b_m+kK_D^{i}\right)s+ k(1+K_P^{i})},\\\nonumber
     ~_\tau\!T_c^{fa}(s) &=  ~_\tau\!T_c^{i}(s),\\\nonumber
     ~_{\ddot{\theta}}\!T_c^{fa}(s) &= \alpha^{fa}j_m Q^{fa}(s)\frac{H_c^{i}(s)}{C^i(s)}
\end{align}
Due to the fact that acceleration measurements generally need to be filtered due to the noisy nature of the signal, a filter~$Q^{fa}(s)$ was added to the signal. In the following, we define the filter to be a second order Butterworth filter as described in Eq.~\ref{eq:butter}. As already described by \citet{pratt1995series}, the acceleration feedback gain~$\alpha^{fa}$ needs to be smaller than one, since an effective reduction to zero inertia would cause instability. A formal reason for this instability again comes from the non-passive apparent impedance a gain of~$\alpha^{fa}=1$ would cause. In the case of an unfiltered acceleration feedback with a PD or FSFt controller, i.e.~$Q^{fa}(s)=1$, the gain is subject to the following constraint to guarantee a passive apparent impedance:
\begin{equation}\label{eq:fbacc_limit}
    \alpha^{fa}\leq \delta_{\zeta}
\end{equation}
In this case, the passivity is violated by a phase lag. In contrast, if the acceleration feedback is filtered, the gain is limited by added phase lead, similar to the effect caused by a DOB. Consequently, the acceleration feedback gain is constrained similarly to the DOB gain, as shown by~\citet{rampeltshammer2020improved}. The exact constraint is given in Appendix~\ref{sec:apdx_1}. 
 
 \begin{figure}
	\centering
    \includegraphics[]{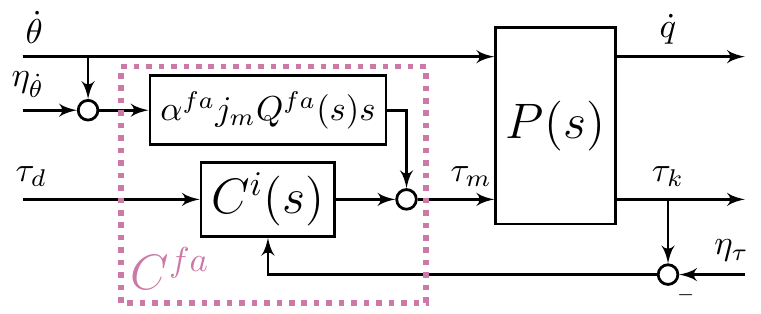}
	\caption{Control scheme for output acceleration feedback with an arbitrary controller~$C^{i}$. The noise for the acceleration signal~$\eta_{\ddot{\theta}} = s\eta_{\dot{\theta}}$ is replaced in the diagram for clarity.}
	\label{fig:acc_fb}
\end{figure}

\section{Nominal system comparison}\label{sec:nominal}
To further analyze the presented state-of-the-art controllers, they are compared based on their closed loop torque transfer, apparent impedance and noise sensitivity in this section. For a numerical example of a SEA, we use the parameters of the SEA developed for the Symbitron exoskeleton~\citep{meijneke2021symbitron, rampeltshammer2020improved}. The tuning  and system parameters used for the subsequent analysis can be found in Table~\ref{table:paramsTheory}.
\begin{table}[h!]
\small\sf\centering
\caption{Tuning parameters for the theoretical comparison of all controllers.}
\label{table:paramsTheory}
 \begin{tabular}{r l} 
 \toprule
 Parameter & Value \\
 \midrule
 Reflected motor inertia $j_m$ & $0.9581\text{ kgm}^2$ \\
 Reflected motor damping $b_m$ & $1.9162\text{ Nms/rad}$ \\
 Spring stiffness $k$ & $1535\text{ Nm/rad}$\\
 Target bandwidth $\omega_{BW}$& $30\text{ Hz}$\\
 Target damping ratio $\zeta_d^{ft,fm,mrac}$ & $0.7$\\
 Target damping ratio $\zeta_d^{ pd}$ & $1$\\
 Torque sensor noise $\sigma_{\tau}$ & $0.0262\text{ Nm}$\\
 Motor velocity sensor noise $\sigma_{\dot{q}}$ & $0.0018\text{ rad/s}$\\
 Output acceleration noise $\sigma_{\ddot{\theta}}$ & $0.2107\text{ rad/s}^2$\\
 \midrule
 \multicolumn{2}{c}{Cascaded PID controller}\\
 \midrule
 $K_{P,o}^{cp}$ & $1\text{ rad/Nms}$\\
 $K_{D,o}^{cp}$ & $0.015\text{ rad/Nm}$\\
 $K_{I,o}^{cp}$ & $0.5\text{ rad/Nms}^2$\\
 $K_{P,i}^{cp}$ & $6\text{ Nms/rad}$\\
 $K_{I,i}^{cp}$ & $3\text{ Nm/rad}$\\
 \midrule
 \multicolumn{2}{c}{DOB variants}\\
 \midrule
 Filter frequency~$\omega_q^{dob}$ & $10\text{ Hz}$\\
 Gain~$\alpha^{dob, pd}$ & $0.646$\\
 Gain~$\alpha^{dob, ft}$ & $0.347$\\
 \midrule
 \multicolumn{2}{c}{Acceleration feedback variants}\\
 \midrule
 Filter frequency~$\omega_q^{fa}$& $20\text{ Hz}$\\
 Gain~$\alpha^{fa, pd}$ & $0.8$\\
 Gain~$\alpha^{fa, ft}$ & $0.45$\\
 \bottomrule
 \end{tabular}
\end{table}

\subsection{Torque loop shaping}
In a first step, it is evaluated, whether the presented controllers can be tuned to show a similar torque tracking. Therefore, each controller was tuned according to the settings provided in Table~\ref{table:paramsTheory}. It has to be noted that the damping ratio of the FSFt and FSFm controllers is different to the damping ratio of the PD controller. While the damping ratio of the PD controller is reducing the magnitude of the resonance peak, it cannot completely eliminate it. This effect is mainly caused by the zero in the transfer function. The cascaded PID controllers was tuned to match the resonance peak of the PD controller, and its limited gains were set to the maximally allowable gains. 
\begin{figure}
	\centering
    \includegraphics[]{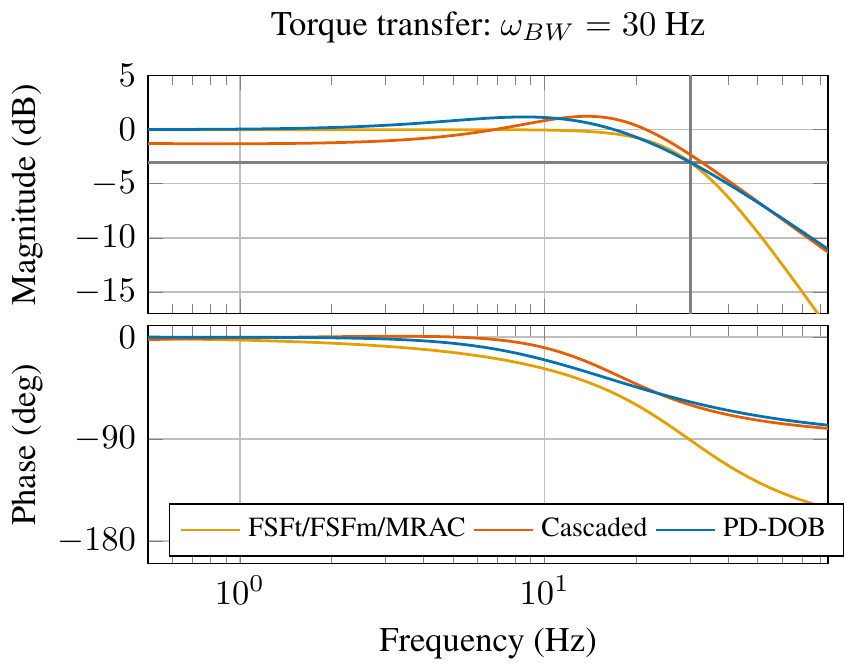}
	\caption{Theoretical comparison of the torque tracking performance of the presented controllers. Controllers are tuned for a bandwidth~$\omega_{BW}=30~\text{Hz}$ and similar resonance peaks. FSFt and FSFm (not shown) have identical torque transfer. Additionally, ideal MRAC (not shown) torque response is identical to FSFt.}
	\label{fig:r1_torTheory}
\end{figure}
In Fig.~\ref{fig:r1_torTheory} the results for the PD, FSFt, FSFm, and Cascaded PID controller are shown. The ideal MRAC torque tracking, i.e. equal to its reference model, is identical to the FSFt response. This demonstrates that the presented approaches can be tuned for a similar torque tracking response. For the Cascaded PID controller tuning is an iterative process and might be best solved by using an optimization process that defines bandwidth and resonance peak heights. The major difference between the controllers is the higher overshoot of PD and Cascaded PID controllers.

\subsection{Apparent impedance}
For the apparent impedance, the FSFt, FSFm, PD, and Cascaded PID controllers are evaluated. Additionally, the DOB and acceleration feedback are evaluated alongside an inner loop PD and FSFt controller. As both MRAC and Cascaded PID have a more complex control law, we did not evaluate DOB or acceleration feedback additions, since the overall passivity limits would be hard to determine. All controllers are tuned for similar torque tracking performance. Additionally, filter bandwidths for DOB and acceleration feedback variants are set identically. Gains are selected such that the apparent impedance is marginally passive, which results in the lowest possible apparent impedance that is still passive. The selected parameters are shown in Table~\ref{table:paramsTheory}.
\begin{figure*}
	\centering
    \includegraphics[]{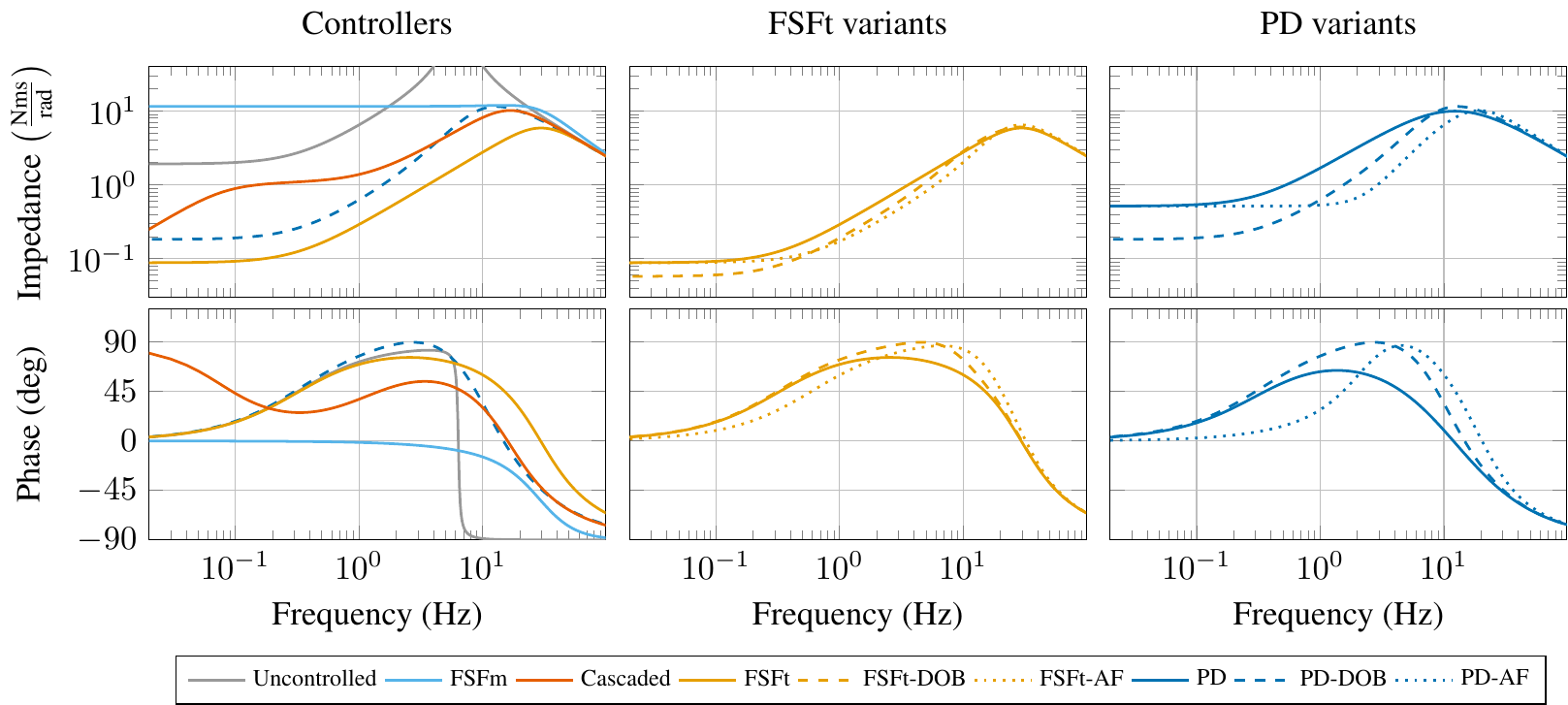}
	\caption{Theoretical comparison of the apparent impedance. On the left side the state of the art controllers PD-DOB, FSFt, FSFm and Cascaded PID are compared. On the right side, both PD and FSFt controllers' apparent impedance is improved with a DOB (dashed lines) and acceleration feedback (dotted lines). All gains are selected to achieve a marginally passive apparent impedance.}
	\label{fig:r2_impTheory}
\end{figure*}

In Fig.~\ref{fig:r2_impTheory}, the comparison between the selected controllers is demonstrated. It can be seen that FSFt shows the theoretically lowest apparent impedance, while FSFm shows the highest apparent impedance at low frequencies. This discrepancy, despite identical torque tracking performance, comes from the difference of information contained in the feedback signals of torque rate~$\dot{\tau}_k$ and motor velocity feedback~$\dot{q}$. The torque rate carries additional information about the load side velocity~$\dot{\theta}$, which lowers the apparent impedance. 

The Cascaded PID controller has the capability to lower the apparent impedance compared to the uncontrolled plant, and even has the lowest apparent impedance for frequencies close to zero, as it tends to zero. However, the increased order of the transfer function can be seen on the worse performance in the range of~${\omega\in[0.1, 1]~\text{Hz}}$, were it performs worse than most controllers. It can also be seen that the phase lead of the Cascaded PID controller is less than FSFt and PD-DOB in that frequency range. Most likely, this can be attributed to the conservative passivity limits proposed by~\citet{vallery2007passive}.

Improving the apparent impedance with a DOB mostly lowers the apparent impedance at lower frequencies, as shown for both PD-DOB and FSFt-DOB variants compared to the apparent impedance of those controllers without DOB. In contrast, acceleration feedback mostly improves the apparent impedance in the medium frequency range~${\omega\in[1, 10]~\text{Hz}}$. Both methods increase the phase lead of the apparent impedance, thus causing non-passivity if their respective gains are not selected correctly. The limit in both cases is the introduced phase lead. From the selected parameters, it can be seen that FSFt variants have lower gains for DOB and acceleration feedback, compared to PD variants. At the same time FSFt by itself has a lower apparent impedance compared to a PD controller. In general, these results indicate that it is only reasonable to select a single method to improve the apparent impedance to achieve a maximum effect in the target frequency range. Otherwise, the effect of the apparent impedance shaping methods would be to small. Additionally, these results indicate a hard limit on lowering the apparent impedance while keeping it passive.

\subsection{Noise sensitivity}
\begin{figure*}
	\centering
    \includegraphics[]{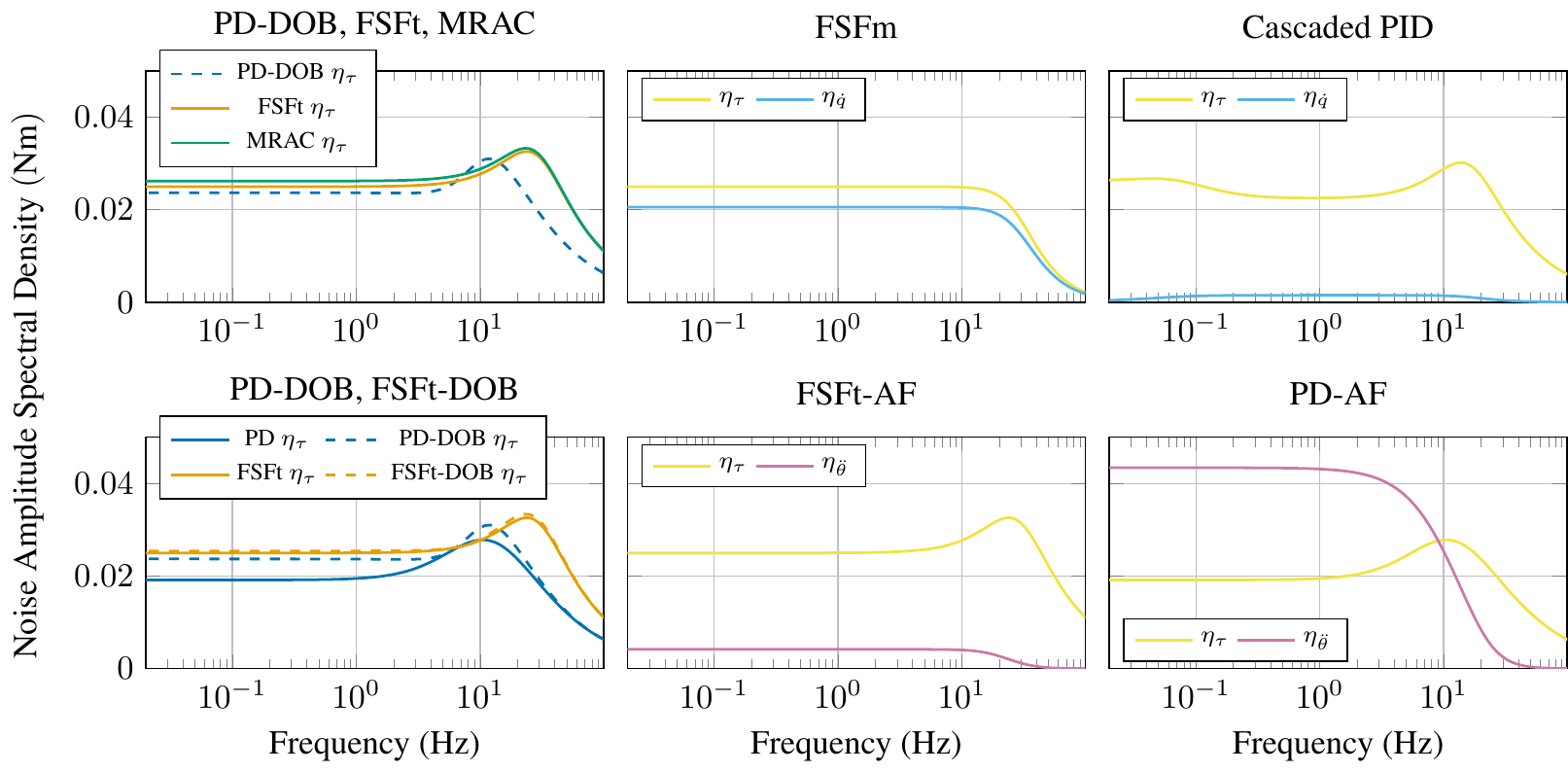}
	\caption{Theoretical comparison of the noise spectral density for each controller. The panels in the left column show all controllers with pure torque feedback~(FSFt, PD, MRAC, FSFt-DOB, PD-DOB). The top panels show the noise sensitivity for all presented torque controller, with the bottom panels showing controllers with apparent impedance shaping methods. The noise spectral density demonstrates the effect of each noise source on the controlled interaction torque~$\tau_k$.}
	\label{fig:r3_noise}
\end{figure*}

The last point to be considered is the noise sensitivity of all controller approaches. For this analysis, all controllers are again tuned based on the parameters shown in Table~\ref{table:paramsTheory}. To effectively compare controllers with different feedback sources, and consequently different noise sources, we decided to look at the noise amplitude spectral density~$N^i_n$, which is defined as the square root of the power spectral density of each noise source~\citep{cerna2000fundamentals}. For a white noise source with a standard deviation of~$\sigma_n$ this results in
\begin{align}
    N^i_n = ~_n\!T^{i}_c \sigma_n.
\end{align}
The noise spectral density is normalized with respect to the noise source, and as such enables us to compare the effects of different noise sources to each other. The resulting amplitude spectra for sensors used in the test setup, standard deviations are shown in Table~\ref{table:paramsTheory}, are shown in Figure~\ref{fig:r3_noise}. This analysis might lead to different results based on the individual SEA and its equipped sensors.

These amplitude spectra show that both FSFt and MRAC are more sensitive to noise compared to a PD controller starting from a frequency~$\omega=10~\text{Hz}$. FSFm generally has a higher noise level due to comparable noise spectra for both of its feedback signals. The cascaded PID controller exhibits similar noise sensitivity to the PD controller, and noise on its internal velocity loop has barely any influence on the controller performance. For DOB variants it can be seen that a DOB slightly increases noise amplitudes at lower frequencies, which is consistent for both FSFt-DOB and PD-DOB. For the acceleration feedback variants, the difference between variants is bigger, since FSFt-AF has a significantly lower contribution of the acceleration sensor, compared to PD-AF. However, the noise spectrum for the torque sensor is significantly lower for the PD variant compared to the FSFt variant. These results indicate that a PD controller and its variants, as well as a Cascaded PID are less influenced by noise in the system for this specific setup.
\section{Experimental Methods}\label{sec:expmethods}
In the following, the experimental setup, the controller implementation, as well as the evaluation protocol are described.

\begin{figure}
	\centering
    \includegraphics[]{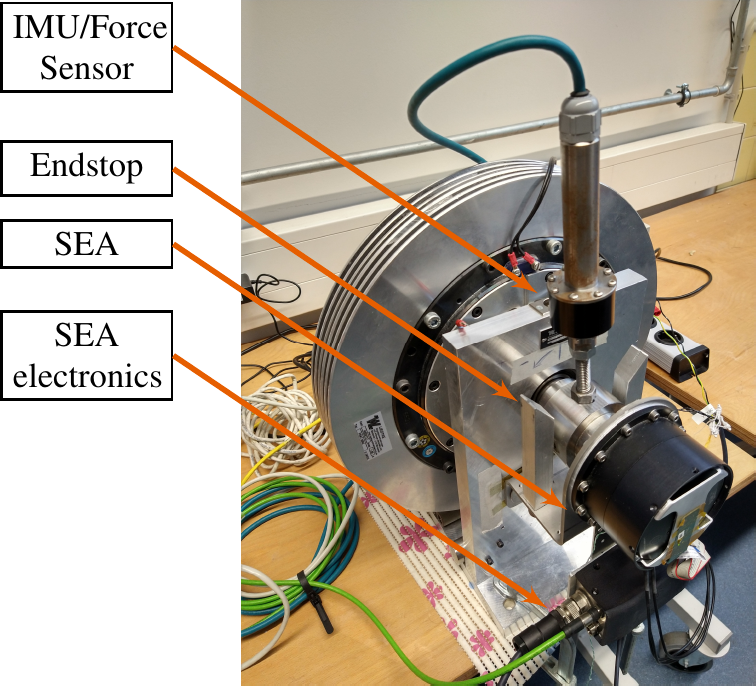}
	\caption{Setup of SEA for experimental evaluation.}
	\label{fig:setup}
\end{figure}

\subsection{Setup}
The controller presented in this work is designed for the SEA of the Symbitron exoskeleton~\citep{meijneke2021symbitron} and tested on one of its actuators as shown in Fig.~\ref{fig:setup}. A Tiger Motor U8-10(Pro), T-Motor, Nancheng, China, is reduced by a LCSG20 harmonic drive, Leader Drive, Jiangsu, China, which connects to a custom rotary spring. The motor is controlled by an Ingenia Everest Net drive, Ingenia-Cat, Barcelona, Spain, and communicates over EtherCAT. Furthermore, the actuator is equipped with two Aksim encoders from RLS~(Renishaw), each with a resolution of 20 bits; one to measure the joint angle, and the other to measure the spring deflection. The motor position is measured by a MHM encoder, IC Haus, with a resolution of 16 bits. The joint angle and spring deflection encoders are connected to the motor drive via a custom daisy chainer and run at $1$ kHz. The actuator is controlled via TwinCAT in a compiled Simulink model.

The load side of the actuator was designed to deliver torques up to 100~Nm, and can achieve speeds up to 5 rad/s. The reflected motor inertia was computed from the known motor inertia and gearbox specifications, and the reflected motor damping was identified from the open loop torque response of the actuator. Spring stiffness of the custom spring was identified from its CAD model, and verified by attaching it to a torque sensor and identifying its stiffness.

The setup allows the output of the actuator to be locked by fixing it with a bracket to the frame of the setup. If not fixed, it can be moved around by a handle that is attached to the output side of the actuator. On top of the handle a BOTA Rokubi force sensor with an integrated IMU, BOTA, Switzerland, is attached to measure the rotational acceleration of the output side. The data from the IMU is read out via EtherCAT at a rate of $800$ Hz.

\subsection{Controller implementation}
All controllers were implemented in Simulink, and compiled for use in TwinCAT. All controllers implemented used the following building blocks:
\begin{itemize}
    \item Torque derivatives are filtered with a second order Butterworth filter with a cutoff frequency of~$160$ Hz.
    \item Motor velocity is unfiltered.
    \item Integrators are implemented as discrete leaky integrators with a linear leak rate of~$a_{leak} = 0.999$ to avoid windup, and to allow identical integrator behavior across different controllers.
    \item Filters and reference models are discretized via the bilinear transform.
\end{itemize}
Additionally to these shared implementations, all other target parameters are shown in Table~\ref{table:paramsExperiment}. This includes the learning parameters for the MRAC, as well as all controller setting for each tested condition. It has to be noted that the target bandwidth for both FSFt and FSFm needed to be reduced to achieve the desired bandwidth. This is most likely caused by small errors in the identified parameters. Conditions with N/A caused oscillations and were eliminated as a result. Additionally, the adaptive torque rate gain~$\hat{b}$ for the MRAC was set to zero for all experiments due to instability it caused.
\begin{table}[h!]
\small\sf\centering
\caption{Tuning parameters for the experimental comparison of all controllers.}
\label{table:paramsExperiment}
 \begin{tabular}{l c c c c} 
 \toprule
 Parameter & $20$ Hz &  $30$ Hz & $40$ Hz & Unit \\
 \midrule
 $\omega_{BW}^{pd}$ & $20$ &  $30$ & $40$ & $\text{Hz}$\\
 $\zeta_d^{pd}$& 1 &  1 & 1 & \\
 $\omega_{BW}^{ft, fm, mrac}$ & $15$ &  $19$ & N/A & $\text{Hz}$\\
 $\zeta_d^{ft, fm, mrac}$& 0.7 &  0.7 & N/A &\\
 \midrule
 $\alpha^{dob, pd}$ & 0.74& 0.65 & N/A&\\
 $\alpha^{dob, ft}$ & 0.44& 0.35 & N/A&\\
 $\alpha^{fa, pd}$ & 0.9& 0.8 & N/A&\\
 $\alpha^{fa, ft}$ & 0.62& 0.45 & N/A&\\
 \midrule
  \multicolumn{5}{c}{Cascaded PID controller}\\
 \midrule
 $K_{P,o}^{cp}$ & \multicolumn{3}{c}{$1$} & $\text{rad/Nms}$\\
 $K_{D,o}^{cp}$ & \multicolumn{3}{c}{$0.015$}& $\text{rad/Nm}$\\
 $K_{I,o}^{cp}$ & \multicolumn{3}{c}{$0.5$}& $\text{rad/Nms}^2$\\
 $K_{P,i}^{cp}$ & $3.75$& $6$& $8$& $\text{Nms/rad}$\\
 $K_{I,i}^{cp}$ & $1.875$ & $3$ & $4$& $\text{Nm/rad}$\\
 \midrule
  \multicolumn{5}{c}{MRAC}\\
  \midrule
 $\omega_{BW}^{r}$ & $20$ &  $30$ & $40$ & $\text{Hz}$\\
 $\hat{b}$ & \multicolumn{3}{c}{$0$} & $\text{s}$ \\
 $\rho$ & \multicolumn{3}{c}{$0.999$} & \\
 $\sigma$ & \multicolumn{3}{c}{$0.001$} & \\
 \bottomrule
 \end{tabular}
\end{table}

\subsection{Identification experiments}
For experimentally comparing all shown controllers, three different tests were conducted: an identification of the torque tracking to demonstrate that similar torque responses can be achieved on an experimental setup and to identify problem sources of the torque tracking, an identification of the apparent impedance to show the apparent impedance and its passivity in the relevant range, and an impact test to demonstrate the passivity of the apparent impedance in a non-linear condition. Data for all experiments was stored at a frequency of $1$ kHz.

For the torque tracking identification, the output of the actuator was fixed to the test setup, and a target sine wave with a single specified frequencies and amplitudes was to be tracked. Individual frequencies were separated by a small pause were zero torque was applied to avoid overheating of the motor when tracking high frequencies. All controllers were to be tested for target bandwidths of $\omega_{BW}\in \left\{20, 30, 40\right\} ~\text{Hz}$. To identify the torque response if each condition, each sine was repeated 15 times with a pause of $0.1$ seconds between each frequency. For the identification 47 frequencies from the range~$\omega \in \left[0.05, 80\right]~\text{Hz}$ were selected, and the resulting sinusoidal excitation signal had amplitudes of $20$ Nm in the range of~$\omega \in \left[0.05, 1\right]~\text{Hz}$, amplitudes of $10$ Nm in the range of~$\omega \in \left[1, 10\right]~\text{Hz}$, and amplitudes of $5$ Nm in the range of~$\omega \in \left[10, 80\right]~\text{Hz}$. For the~$40$ Hz conditions, amplitudes of $2.5$ Nm in the range of $\omega \in \left[16, 80\right]~\text{Hz}$ were applied for all controllers besides MRAC which needed amplitudes of $1$ Nm. The overall decrease in amplitude with frequency is necessary to not saturate the motor current, and thus guarantee accurate tracking~\citep{lee2021performance}. The~$40$ Hz controllers showed strong oscillations for the FSFt, FSFm and MRAC controller in the pauses between excitations, and were not included in the analysis as a result. Torque tracking was evaluated by computing the fast fourier transform of the unfiltered interaction torque and the excitation signal over the full excitation period and dividing them at the relevant frequency. 

\begin{figure*}
	\centering
    \includegraphics[]{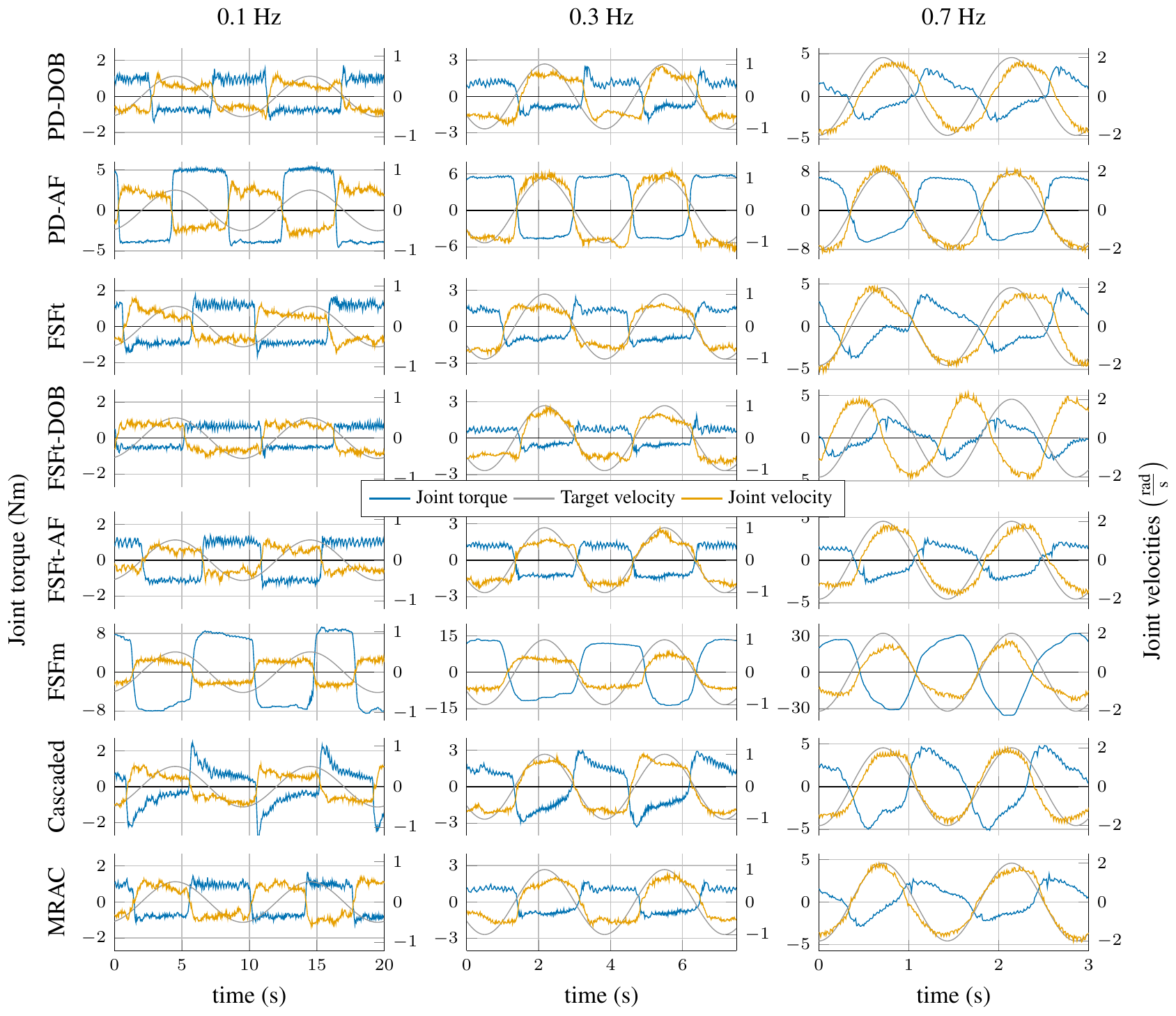}
	\caption{Time series of the selected excitation frequencies~$\omega\in\left\{0.1, 0.3, 0.7\right\}~\text{Hz}$ of the apparent impedance identification experiment for the target bandwidth of~$20$ Hz. Each column has identically scaled axes, with the exception of the PD-AF and FSFm controllers, which exhibit higher torque responses. The data show that the target frequency and amplitude were reached consistently. It can also be seen that the excitation signal at low frequencies is more rectangular, compared to higher frequencies. Additionally, differences in apparent impedance magnitude can be observed from differences in the shape and magnitude of the joint torque.}
	\label{fig:r5a_impIDtimseries}
\end{figure*}

For the apparent impedance identification, the output of the actuator was movable, and the excitation signal was generated manually by a human tester. All controllers were evaluated for for target bandwidths of $\omega_{BW}\in \left\{20, 30\right\} ~\text{Hz}$. The $40$ Hz condition was omitted, due to the oscillatory behavior of the FSFt, FSFm and MRAC controllers in that condition. To identify the apparent impedance, the human user was instructed to follow a reference profile of the output velocity that was shown to them on a screen as accurately as possible, while the torque controller controlled the interaction torque at zero, as shown in Fig.~\ref{fig:r5a_impIDtimseries}. The reference profile consisted of a sine wave of a selected frequency that was repeated 15 times, or at least for 5 seconds, whichever was longer. The apparent impedance was identified at frequencies~$\omega\in\left\{0.1, 0.15, 0.2, 0.3,..., 1, 1.5, 2, 3, ..., 10\right\}$ Hz. The amplitude of the excitation signal was given as $0.5$ rad/s for $\omega\in[0.1, 0.15]$, as $1$ rad/s for $\omega\in[0.2, 0.3]$, as $1.5$ rad/s for $\omega\in[0.4, 0.6]$ Hz, as $2$ rad/s for $\omega\in[0.7, 1]$ Hz,  as $2.5$ rad/s for $\omega\in[1.5, 2]$ Hz, and as $3$ rad/s for $\omega\in[3, 10]$ Hz. The amplitudes were determined based on the preferences of the tester. Apparent impedance was evaluated by computing the ratio of the fast fourier transform of the unfiltered interaction torque and the unfiltered output velocity at the excited frequencies over the last 10 of the 15 repetitions. With the excitation signal being generated by a human, the input signal was not perfectly sinusoidal at low frequencies, as shown in Fig.~\ref{fig:r5a_impIDtimseries}, where the achieved excitation signals at low frequencies are more rectangular. This divergence in shape at low frequencies can be explained by the higher impact of stiction and Coloumb friction at low velocities, as well as the limited range of motion of the test setup, as shown in Fig:~\ref{fig:setup}. The imperfect sinusoidal human input motion does not contribute a visible effect in the division of the frequency spectra, since the enforced motion and generated torques were not saturating motor currents, such that these non-linearities do not occur. Furthermore, such an imperfect input motion also does not have any theoretical effect on the identification, since we only use the frequency component with the highest excitation.

For the impact tests, the output of the actuator was movable, and a rigid endstop was added on both sides of the actuator to offer an impact object. For the test the output was sped up to a velocity of~$4$ rad/s until a distance of $0.04$ rad from the rigid endstop. After that point, the target torque was set to zero for the impact. Each controller was tuned for the $\omega_{BW} = 30~\text{Hz}$ condition, and the impact test was repeated three times per controller. All controllers besides FSFm approximately reached the target speed of~$4$ rad/s. For processing one of three impacts was selected: All impacts of the same controller showed consistent behavior after rebounding from the endstop. However, due to measurements errors caused by the impact, it was not possible to take an average without eliminating these errors or having them influence the mean. Impacts were centered around the second artifact, and normalized based on the mean of the first 50 samples of the selected impact.

%For processing, joint velocity was filtered with a second order zero phase low pass filter with a cutoff frequency of~$100$ Hz. Data was segmented, velocity was normalized to the velocity at $0.05$ s before impact, and the mean response was taken for all three repetitions.
\section{Experimental results}\label{sec:results}
The results for the torque tracking identification, the apparent impedance identification and the impact test are shown in this section.

\begin{figure*}
	\centering
    \includegraphics[]{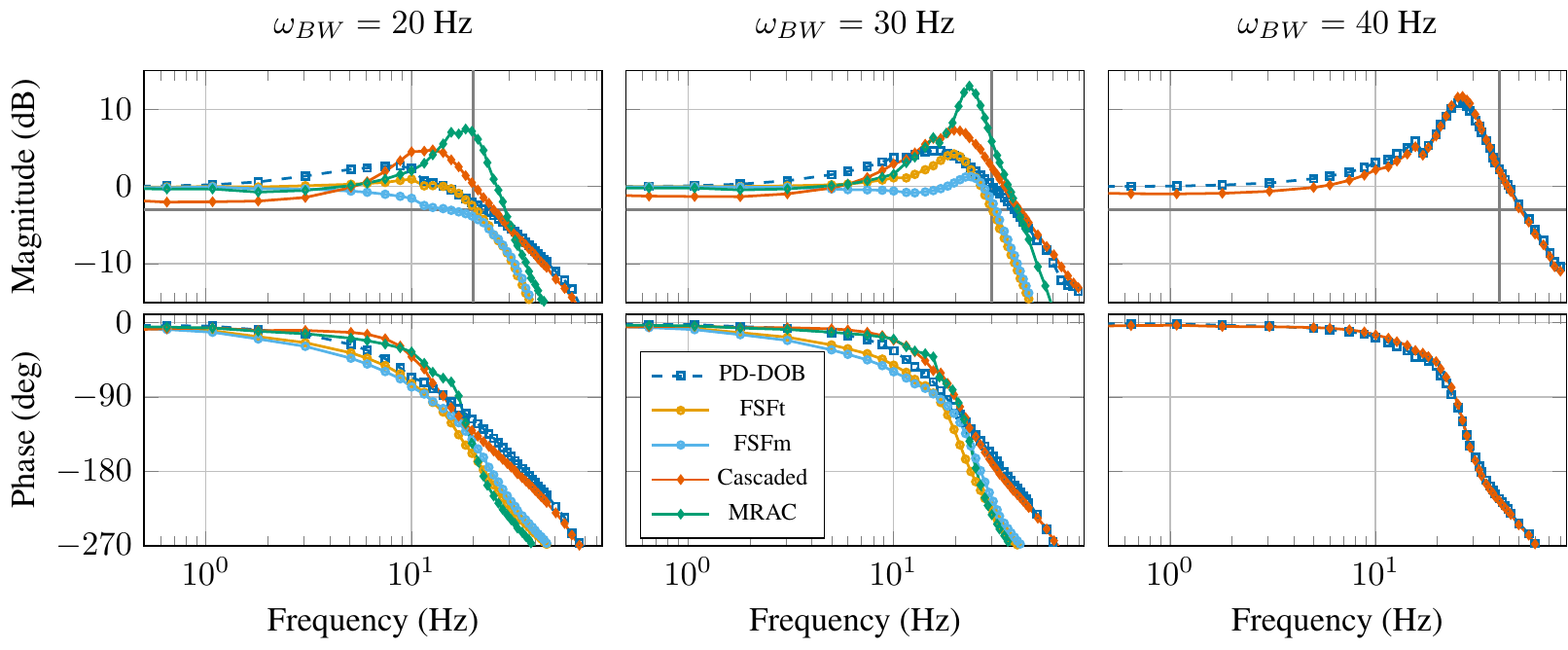}
	\caption{Torque tracking identification results for PD-DOB, FSFt, FSFm, Cascaded PID and MRAC controller. Identification was conducted over a set of frequencies, and each frequency resulted in a data point, as shown in the identified bode plots. Additionally, the target bandwidth is shown with the gray lines.}
	\label{fig:r4_torID}
\end{figure*}

The resulting bode plots of the torque tracking identification are shown in Fig.~\ref{fig:r4_torID}. Is shows that all controllers approximately achieve the desired bandwidths. All controllers show a slightly higher actual bandwidth compared to the target bandwidth, with the strongest effect visible for the MRAC at the $20$ Hz condition. Additionally, both PD-DOB and Cacaded PID controllers show an increase of bandwidth in the $40$ Hz condition. Additionally, the Cascaded PID controller shows imperfect torque tracking at low frequencies. This can most likely be attributed to a slight harmonic distortion and the use of leaky integrators. However, a slight decrease in tracking performance is theoretically expected. 

For all controllers a relative increase in the resonance peak can be seen with increasing bandwidth, even though the peak should stay constant throughout all conditions. As expected from the theoretical results~( see Fig.\ref{fig:r1_torTheory}), both FSFt and FSFm show the lowest resonance peak. The phase of all controllers behaves as expected with an additional phase drop at higher frequencies, caused by the communication delay of the system. The only controllers that were functional at the 40 Hz condition were PD-DOB and the Cascaded PID, with all others exhibiting high frequency oscillations between excitations at this condition, i.e. these controllers were practically unusable at these conditions due to the high noise sensitivity these controllers have at higher frequencies (see Fig.\ref{fig:r3_noise}). Additionally, FSFt, FSFm, and MRAC needed lower target bandwidths to achieve the desired bandwidths. This overall error in bandwidths can be attributed to system parameter identification errors. We only adapted FSFt, FSFm, and MRAC due the stronger bandwidth increase of those methods, compared to PD-DOB and Cascaded PID.

\begin{figure*}
	\centering
    \includegraphics[]{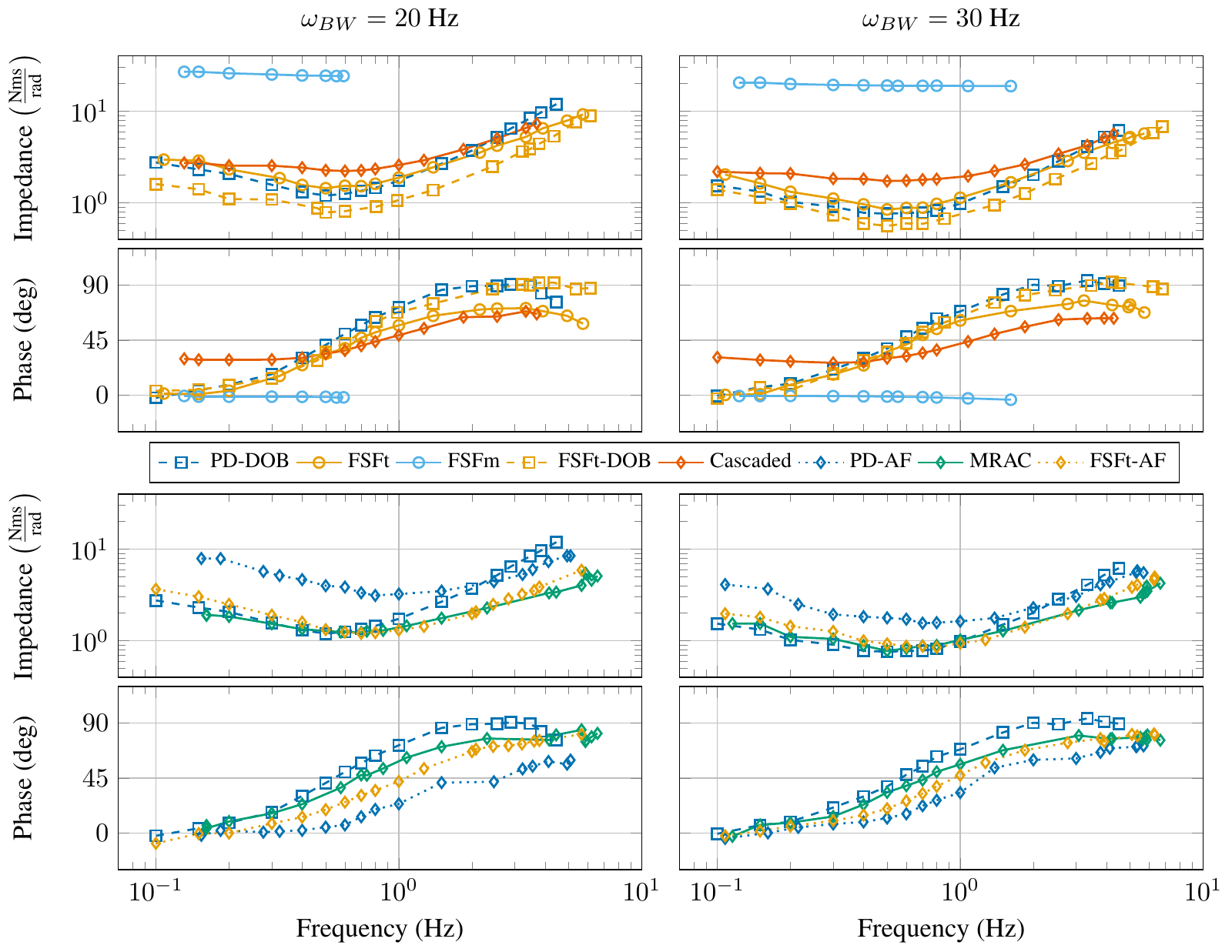}
	\caption{Apparent impedance identification results for PD-DOB, FSFt, FSFm, Cascaded PID and MRAC controller, as well as the improvement method FSFt-AF. Identification was conducted by manually exciting the output of the actuator. From the data points it can be seen that this was achieved consistently. The shown data confirms that apparent impedance can be lowered for target frequency ranges, and that the use of improvement methods causes additional phase lead in the closed loop system.}
	\label{fig:r5_impID}
\end{figure*}

Fig.~\ref{fig:r5_impID} shows the results of the apparent impedance identification. As expected from the theoretical results, the apparent impedance of FSFm is the highest of all approaches. In contrast to the theoretical results FSFt and PD-DOB show a similar apparent impedance. This can most likely be attributed to the increased bandwidth of PD-DOB compared to FSFt. The Cascaded PID controller does not show a lower apparent impedance at low frequencies, as would be expected, and overall has a slightly higher apparent impedance than FSFt and PD-DOB in the range between $\omega\in[0.2, 1]$ Hz, as shown in the theoretical results~(Fig.\ref{fig:r2_impTheory}). This difference, also clearly visible in the time series data in Fig.~\ref{fig:r5a_impIDtimseries}, gets amplified in the higher bandwidth condition. The difference at low frequencies can most likely be attributed to the dominant effect of stiction and dry friction at low frequencies and speeds, as well as the leaky integrators used in the controller implementation. For MRAC, an lower apparent impedance at frequencies above~$\omega=1$ Hz can be observed, behaving similar to FSFt-AF. The jumping of data points at $\omega=5$ Hz is due to instability of the MRAC at these frequencies. FSFt-AF behaves as expected by lowering the apparent impedance at frequencies above $\omega=1$ Hz, and by increasing the phase lead compared to FSFt. All methods are passive, i.e. phase lead lower than $90$ degrees. Both impedance manipulation methods, DOB and acceleration feedback, reach the maximum phase lead for passivity, as expected. All methods have a passive apparent impedance as expected. Additionally, all methods perform worse at lower frequencies, compared to theoretical expectations. This can most likely be attributed to non-linear effects such as stiction and coloumb friction. Overall, FSFt, MRAC and PD-DOB perform similarly, with FSFt-AF and MRAC having reduced apparent impedance at higher frequencies.

\begin{figure*}
	\centering
    \includegraphics[]{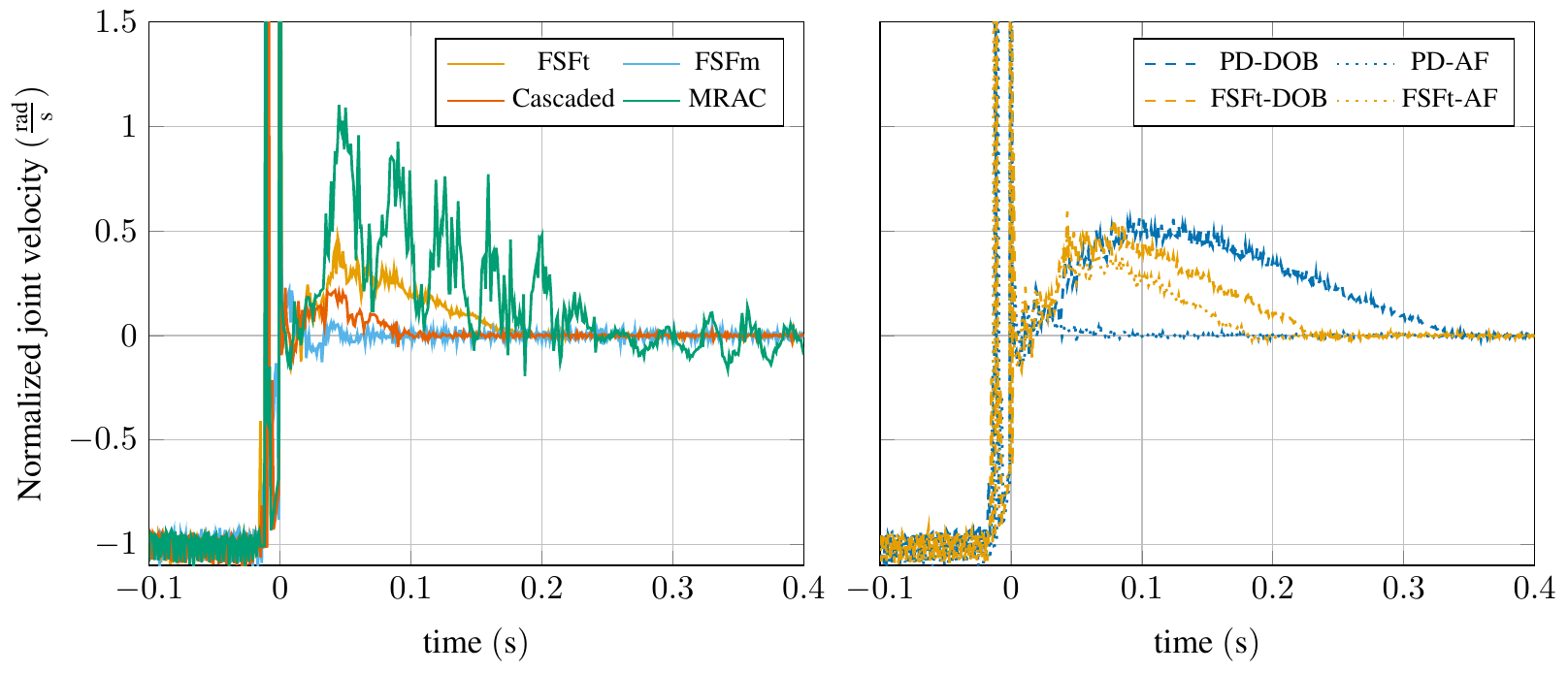}
	\caption{Impact tests for all proposed controllers. Impact happens at the first jump of velocity. In between the first and second jump, the endstop is bending slightly, and rebound happens at time zero. Both jumps of velocity during impact are single sample measurement errors from the encoders. It is shown that all controllers are indeed passive, i.e. all velocities go back to zero at varying speeds, with only MRAC showing oscillations after impact.}
	\label{fig:r6_impact}
\end{figure*}

In Fig.~\ref{fig:r6_impact}, the results for the impact tests are shown. For all controllers, two measurement errors can be observed, both of which are present for exactly one sample. The first error coincides with initial contact with the endstop, while the second error coincides with the start of the rebound from the endstop. In between both artifacts the endstop deforms due to the high torques (up to 70 Nm) generated by the impact. All controllers are passive, i.e. their peak velocity after impact is always lower than peak velocity before impact. Main difference between all controllers is the time it takes the controller to stop the actuator, i.e. $\dot{\theta} = 0$, which is lowest for FSFm and Cascaded PID, and highest for PD-DOB. MRAC showed the biggest demerit with an oscillation after impact.

\section{Discussion}\label{sec:discussion}
In the following, we are discussing controller performance based on theoretical and experimental results in the presented unified framework. The discussion is mainly focused on showing the relative merits and demerits of each presented controller to aid in the choice for controller that is appropriate for a specific actuator or application.

From the torque tracking results, it can be concluded, that all proposed methods achieve similar torque tracking performance. Major differences are that FSFt and FSFm achieve a lower resonance peak, while PD-DOB and Cascaded PID can achieve higher bandwidths without showing oscillating behavior. The former is theoretically expected, while the latter can be attributed to the lower noise sensitivity achieved by PD-DOB and Cascaded PID controllers compared to the FSF variants or to the fact that the controller gains negatively influence the maximum torque a SEA controller can track simply by their controller gains, which is a persistent problem of SEAs~\citep{lee2021performance}. Additionally, it can be concluded that FSF variants are more sensitive to system parameter errors, since it was necessary to adapt the target bandwidth to achieve the desired bandwidths.

The apparent impedance identification demonstrated that FSFt, MRAC, and PD-DOB can achieve a similar apparent impedance at low frequencies. For FSFt this result is worse than theoretically expected, but this difference can be explained by errors of the system parameters and as a result a lower target bandwidth which results in lower gains. The cascaded PID controller shows comparable apparent impedance to the best three approaches. It is likely that its performance can be further improved by utilizing loop shaping and relaxing its conservative passivity constraints by evaluating the full passivity condition instead of enforcing positive values for all polynomial coefficients as done for the PD-DOB approach in our previous work~\citep{rampeltshammer2020improved}. This possible improvement can also be seen in the shape difference of the torque response in Fig.~\ref{fig:r5a_impIDtimseries}. MRAC showed the best performance at low frequencies but tended to start oscillating at higher frequencies. High performance at low frequencies was achieved by high forgetting factor, i.e. slow adaptation. However, this slow adaptation becomes a problem at higher frequencies causing the observed oscillations. The resulting trade-off, i.e. faster adaptation, will cause a worse performance at low frequencies at the benefit of general usability, has to be considered when analyzing and using this adaptive controller. Additionally, it could only partially adapt, since the damping adaptation gain was unstable for the used setup. Another problem of MRAC is the non-linearity of its adaptation, which introduces non-linear noise problems, i.e. the reason the damping adaptation could not be used, and can cause (partial) instability, as shown with the oscillatory response during impact tests. This clearly limits its use for applications that need guaranteed performance.

Apparent impedance improvement methods performed as expected, with the DOB lowering the apparent impedance at low frequencies, and acceleration feedback lowering the apparent impedance at higher frequencies. Additionally, experimental results confirmed the increased phase lead for the use of DOB or filtered acceleration feedback. Due to this increase in phase lead, it is obvious that there is limited room to improve apparent impedance for a closed loop system. Hence, it is advisable to select an improvement method based on the application case  with its respective requirements at high or low frequencies.
 
The impact tests confirmed that all presented controllers are indeed passive, and also highlighted some small differences across controllers based on the time it took for the actuator to come to rest after each impact, i.e.~$\dot{\theta}=0$. Controllers with a generally lower apparent impedance~(PD-DOB, FSFt-DOB, MRAC) took longer to come to rest compared to controllers with a generally higher apparent impedance~(FSFm, PD-AF) stopped almost immediately after impact. This behavior is expected based on the more mass-like behavior controllers with low apparent impedance show.

From the tested state of the art controllers, there exists no clear best option. FSFt is a simple controller, with a low apparent impedance at low frequencies and easy to tune, but is quite sensitive to noise. FSFm is not recommended as a torque controller for SEAs due to its high apparent impedance and noise sensitivity. PD-DOB is more complex to tune compared to FSFt, but can achieve a higher bandwidth and is less sensitive to sensor noise compared to FSFt. Theoretically, it has a higher apparent impedance compared to FSFt, but experimentally, the difference seems non-existent. Cascaded PID performs comparable to PD-DOB, but suffers from a worse torque tracking at low frequencies in ourt test bench, and is more complex to tune. Additionally, the conservative passivity limits used by \cite{vallery2007passive} most likely increase the apparent impedance at low frequencies. By evaluating the whole passivity condition for positive realness, instead of its polynomial coefficients, the controller can most likely achieve a lower apparent impedance while maintaining interaction stability. MRAC has good torque tracking performance, and low apparent impedance, but suffers from its complexity. The additional adaptation helps its performance but is hard to tune over a broad frequency range and can also easily get unstable. Additionally, its non-linearity introduces additional noise amplification, e.g. torque rate adaptation gain is not usable, and does not guarantee overall stability. This lack of robustness makes its use in complex environments a risk for the system.

As a conclusion, it can be seen that these controllers have a trade-off between low noise sensitivity and lower apparent impedance, when achieving a comparable torque tracking.

By analyzing the apparent impedance of SEA controllers instead of analyzing their interaction with specific linear environments~\citep{calanca2018rationale, paine2015actuator}, specific limits on interaction stability can be found. For example, the limits for acceleration feedback proposed by~\cite{pratt1995series} and~\cite{calanca2018rationale} are solely based on parameter estimation errors and not on the necessary filtering of acceleration signals. Additionally, looking at the apparent impedance helps to identify, how the controlled actuator performs over a broad range of conditions by separating the detrimental effects, i.e. torque tracking error cause by external motion, from the torque tracking bode plots, as commonly used in the field~ \citep{calanca2018rationale, paine2015actuator}. Instead of assessing controlled actuators for a specified second order environment, looking at the apparent impedance can give insights into possible errors across frequency regions, thus generalizing results across environments. Based on this reasoning, a DOB is not designed to improve torque tracking near the systems bandwidth, while acceleration feedback is useful in those regions.
\section{Conclusion}\label{sec:conclusion}
In this work, we analyzed state of the art controllers and new combinations of controllers for SEAs based on a unified framework by analyzing theoretical and experimental torque tracking performance, apparent impedance and noise sensitivity. Based on this analysis, the presented state of the art controllers have specific advantages and disadvantages by trading of lower apparent impedance, i.e. less tracking error introduced by external motions, for a higher noise sensitivity, and vice versa. These trade-offs could be investigated easily by utilizing the apparent impedance as an analysis tool instead of simulating a set of interactions with fixed environments.
Additionally, DOB and acceleration feedback were analyzed as methods to improve the apparent impedance. From the analysis it can be concluded that a DOB lowers the apparent impedance at lower frequencies, thus being helpful for static interaction scenarios, e.g. a human-robot interaction, while the effect of acceleration feedback is more obvious at higher frequencies, indicating its benefits for tasks at higher frequencies. 
In future work, the passivity condition of the Cascaded PID controller should be further investigated and loop shaping methods should be utilized to improve its performance. Additionally, non-linear adaptive controllers such as the MRAC need to be further investigated to see, if the existing limit on lowering the apparent impedance can be reduced while keeping the system's apparent impedance passive. Therefore, a general stability analysis of MRACs in non-linear interaction scenarios also should be investigated.

\begin{acks}
The author's would like to thank Quint Meinders for the design and manufacturing of the test setup, and Michiel Ligtenberg for maintenance and adaptation of the test setup.
\end{acks}
\begin{dci}
The authors declare that there is no conflict of interest.
\end{dci}
\begin{funding}
This research is part of the Flexible Robotic Suit programme by the Dutch Organisation for Scientific Research (NWO), under grant number 14429.
\end{funding}

\bibliographystyle{SageH}

\appendix
\section{Appendices}\label{sec:apdx_1}
\subsection{Derivation of control laws}
This section derives changes in notation and re-arrangement of terms for specified control laws.
\subsubsection{FSFBm}
Equivalence of control laws shown in this work and the one presented by~\citet{losey2016time}:
\begin{align}
    \tau_m &= -\left(j_m\omega_d^2-k\right)q -K_D^{fm}\dot{q} -K\theta + j_m\omega_d^2q_d\\\nonumber
    &= -a_1q -K_D^{fm}\dot{q} -k\theta + j_m\omega_d^2q_d\\\nonumber
    &= -a_1q -K_D^{fm}\dot{q} -k\theta + j_m\omega_d^2q_d + kq_d-kq_d\\\nonumber
    &= -a_1q -K_D^{fm}\dot{q} +k\left(q-\theta\right) + a_1q_d\\\nonumber
    &= a_1\left(q_d -q\right) -K_D^{fm}\dot{q} + \tau_d\\\nonumber
    &= \left(j_m\omega_d^2-k\right)\frac{k}{k}\left(q_d -q +\theta -\theta\right) -K_D^{fm}\dot{q} + \tau_d\\\nonumber
    &=\tau_d + K_P^{fm}\left(\tau_d-\tau_k\right) - K_D^{fm}\dot{q}
\end{align}

\subsubsection{MRAC}
Equivalence of control laws shown in this work and the one presented by~\citet{calanca2018understanding}, adapted from their application specifications, and resulting transfer function. To achieve similarity of the presented control laws, we define $\lambda_1 = 2\zeta_d\omega_d$, and $\lambda_1 =\omega_d^2$. The same parametrization is used by~\citet{calanca2018understanding} for their experimental section
Hence, the reference model defined as
\begin{align}
    \ddot{\tau}_r& + \lambda_1\dot{\tau}_r + \lambda_2\tau_r = \lambda_2\tau_d\\\nonumber
    \ddot{\tau}_r& + 2\zeta_d\omega_d\dot{\tau}_r + \omega_d^2\tau_r = \omega_d^2\tau_d.
\end{align}
Similarly, we define $\lambda = \omega_d$ for the control input, and with $e=\tau_r-\tau_k$ (Please be aware that we switched $\tau_d$, and $\tau_r$ in this paper compared to Calanca's to align with our standard notation), resulting in
\begin{align}
    \tau_m &= \frac{j_m}{b_m}\left(\ddot{\tau}_r-2\omega_d\dot{e}-\omega_d^2e\right)+\hat{b}'\dot{\tau}_k + \hat{c}\tau_k\\\nonumber
    &= \frac{j_m}{b_m}\left(\omega_d^2\left(\tau_d-e-\tau_r\right) -2\omega_d\left(\zeta_d\dot{\tau}_r+\dot{e}\right)\right)\\\nonumber
    &\quad+\hat{b}'\dot{\tau}_k + \hat{c}\tau_k\\\nonumber
    &= \frac{j_m}{b_m}\left(\omega_d^2\left(\tau_d-\tau_k\right) -2\omega_d\tau_k\right)\\\nonumber
    &\quad+\hat{b}'\dot{\tau}_k + \hat{c}\tau_k+ \frac{2\omega_dj_m}{k}\left(1-\zeta_d\right)\dot{\tau}_r\\\nonumber
\end{align}
Due to differences in the physical SEA model, we replace $\hat{b}'= \frac{2\zeta_n}{\omega_n} + \hat{b}$ to ensure that steady state optimum value for $\hat{b}$ is zero, identically to the model of~\cite{calanca2018rationale}. Additionally, we want to remark that the case of $\zeta_d\neq 1$ is not covered in Calanca's paper, thus avoiding the pole zero cancellation observed in this work.

\subsection{Derivations of passivity limits}
This Appendix contains derivations of passivity limits that were not presented in previous works.
\subsubsection{Disturbance observer}
The passivity limit for a disturbance observer is taken from~\citet{rampeltshammer2020improved}, and can be used for both FSFt and PD variants:
\begin{align}\label{eq:alphamax}
    \alpha^{dob}_{max}=&  \min_\omega \alpha^{dob}(\omega), \forall \omega:  \alpha^{dob}(\omega) > 0\\\nonumber
	\alpha^{dob}(\omega) =& \left(1+\frac{\omega^4}{\omega_q^4} \right)\left(1 - \delta_\zeta + \frac{\omega^2}{\omega_d^2}\delta_\zeta\right)\\\nonumber
	&\Bigg(\left(1 + \frac{\omega^2}{\omega_q^2}\right)\left(1-\delta_\zeta + \frac{\omega^2}{\omega_d^2}\delta_\zeta\right) \\\nonumber
	&+  2\sqrt{2}\zeta_d\frac{\omega^2}{\omega_d\omega_q}\left(\frac{1- \frac{\omega^2}{\omega_d^2}}{4\zeta_d^2} - 1 + \delta_\zeta\right)\Bigg)^{-1},\\\nonumber
\end{align}
with $\delta_\zeta$ being the damping correction factor~(see Eq.\ref{eq:delta}), $\omega_q = \omega_q^{dob}$ the DOB filter bandwidth.

\subsubsection{Acceleration feedback}
Passivity of the apparent impedance is established with the positive real condition, i.e.
\begin{equation}
    Z^{fa}\left(j\omega\right) + Z^{fa}\left(-j\omega\right)\geq 0,\;\forall\omega\in\mathbb{R}
\end{equation}
The condition can be simplified to 
\begin{align}
    0\leq& \left(j\left(1-\alpha^{fa}\right)j_m\omega + b_m\right)\\\nonumber &\left(-j_m\omega^2+k\left(1+K_P^{i}\right) + j\left(b_m+kK_D^{i}\right)\omega\right) \\\nonumber
    & + \left(j\left(1-{\alpha}^{fa}\right)j_m\omega + b_m\right)\\\nonumber
    &\left(-j_m\omega^2+k\left(1+K_P^{i}\right) + j\left(b_m+kK_D^{i}\right)\omega\right)\\\nonumber
    0\leq& 2b_m\left(k\left(1+K_P^{i}\right)-j_m\omega^2\right) \\\nonumber
    &+ 2\left(b_m+kK_D\right)\left(1-{\alpha}^{fa}\right)j_m\omega^2
\end{align}
This results in
\begin{align}
    \alpha^{fa} \leq 1 + \frac{b_mk\left(1+K_P^{i}\right)}{\left(b_m+kK_D^{i}\right)j_m\omega^2}- \frac{b_m}{b_m+kK_D^{i}},
\end{align}
which, in case of $\omega\to\infty$, gets to its global maximum value as shown in Eq.~\ref{eq:fbacc_limit}.

\end{document}